\documentclass[a4paper,11pt]{article}
\pdfoutput=1
\usepackage[utf8]{inputenc}
\usepackage{amssymb}
\usepackage{amsmath}
\usepackage{amsthm}
\usepackage{amsfonts}
\usepackage[pdftex]{graphicx}
\usepackage{geometry}
\usepackage{braket}
\usepackage{enumerate}
\usepackage{comment}

\usepackage{subcaption}

\allowdisplaybreaks

\usepackage[table,usenames,dvipsnames]{xcolor}

\usepackage{setspace}
\usepackage[in]{fullpage}
\usepackage{hyperref}
\usepackage{array}
\usepackage{cancel}
\usepackage{wrapfig}
\usepackage[font=small,labelfont=bf]{caption}
\usepackage{pifont}

\usepackage[dvipsnames]{xcolor}
\usepackage{mathtools}

\usepackage{color}

\usepackage{tikz}
\usetikzlibrary{arrows,shapes}
\usetikzlibrary{trees}
\usetikzlibrary{matrix,arrows} 
\usetikzlibrary{positioning}
\usetikzlibrary{calc,through}
\usetikzlibrary{decorations.pathreplacing}
\usepackage{pgffor}	

\numberwithin{equation}{section}

\usepackage[english]{babel}

\usepackage{caption}

\usepackage[nottoc,notlot,notlof]{tocbibind}
\usepackage[nosort]{cite}   
\usepackage{color}

\usepackage{parskip}
\usepackage[T1]{fontenc}

\usepackage{lmodern}
\usepackage{yfonts}

\usepackage{physics}
\usepackage{slashed}
\usepackage{makeidx}
\usepackage[vcentermath]{youngtab}
\usepackage{amsthm}
\usepackage[scr=boondoxo]{mathalfa}

\hypersetup{
	colorlinks=true,
	linktoc=page,
	citecolor=blue,
	linkcolor=blue,
	urlcolor=blue} 

\usepackage{tikz}
\usetikzlibrary{arrows,shapes}
\usetikzlibrary{trees}
\usetikzlibrary{matrix,arrows} 				
\usetikzlibrary{positioning}				
\usetikzlibrary{calc,through}				
\usetikzlibrary{decorations.pathreplacing}  
\usepackage{pgffor}							

\usetikzlibrary{decorations.pathmorphing}	
\usetikzlibrary{decorations.markings}
\tikzset{
   vector/.style={decorate, decoration={snake, amplitude=1pt, segment length=6pt}, draw,double},
   vector2/.style={decorate, decoration={snake, amplitude=1pt, segment length=6pt}, draw},
	provector/.style={decorate, decoration={snake,amplitude=2.5pt}, draw},
	antivector/.style={decorate, decoration={snake,amplitude=-2.5pt}, draw},
    fermion/.style={draw=black, postaction={decorate},
        decoration={markings,mark=at position .55 with {\arrow[draw=black]{>}}}},
    fermionbar/.style={draw=black, postaction={decorate},
        decoration={markings,mark=at position .55 with {\arrow[draw=black]{<}}}},
    fermionnoarrow/.style={draw=black},
    gluon/.style={decorate, draw=black,
        decoration={coil,amplitude=4pt, segment length=5pt}},
    scalar/.style={dashed,draw=black, postaction={decorate},
        decoration={markings,mark=at position .55 with {\arrow[draw=black]{>}}}},
    scalarbar/.style={dashed,draw=black, postaction={decorate},
        decoration={markings,mark=at position .55 with {\arrow[draw=black]{<}}}},
    scalarnoarrow/.style={dashed,draw=black},
    electron/.style={draw=black, postaction={decorate},
        decoration={markings,mark=at position .55 with {\arrow[draw=black]{>}}}},
	bigvector/.style={decorate, decoration={snake,amplitude=4pt}, draw},
}
\tikzset{cross/.style={cross out, draw, 
         minimum size=2*(#1-\pgflinewidth), 
         inner sep=0pt, outer sep=0pt}}

\tikzstyle{block} = [draw, rectangle, 
    minimum height=3em, minimum width=6em]

\newcommand{\agl}[2]{\langle#1 #2 \rangle}

\newcommand{\lu}[1]{\lambda^{#1}}
\newcommand{\ltu}[1]{\tilde{\lambda}^{\dot{#1}}}
\newcommand{\ld}[1]{\lambda_{#1}}
\newcommand{\ltd}[1]{\tilde{\lambda}_{\dot{#1}}}

\newcommand{\cA}{\mathcal{A}}
\newcommand{\cN}{\mathcal{N}}

\usepackage{float}
\usepackage{marvosym}
\usepackage{empheq}
\usepackage{bbold}

\usepackage{tikz}
\usetikzlibrary{shapes}
\usetikzlibrary{arrows}
\usetikzlibrary{positioning}
\usetikzlibrary{decorations.pathmorphing}
\usetikzlibrary{decorations.pathreplacing}
\usetikzlibrary{decorations.markings}

 \def\cA{\mathcal{A}}
 \def\uno{\mbox{1 \kern-.59em {\rm l}}}

\newcommand{\zb}{\bar{z}}
\newcommand{\partdiv}[2]{\frac{\partial #1}{\partial #2}}
\newcommand{\celSp}{\xi}
\DeclareMathOperator{\cs2}{\mathcal{CS}^2}
\DeclareMathOperator{\cJ}{\mathcal{J}}

\DeclareMathOperator{\sl2c}{\text{SL}(2,\mathbb{C})}
\DeclareMathOperator{\SU4}{\text{SU}(4)}

\newcommand{\cb}{\bar{c}}
\newcommand{\db}{\bar{d}}
\newcommand{\hb}{\bar{h}}
\newcommand{\partialb}{\bar{\partial}}
\newcommand{\ub}{\bar{u}}
\newcommand{\pld}[1]{\partial_{#1}}
\newcommand{\pltd}[1]{\tilde{\partial}_{\dot{#1}}}
\theoremstyle{definition}

\begin{document}


\begin{flushright}
	QMUL-PH-21-23\\
	SAGEX-21-09\\
\end{flushright}

\vspace{20pt} 

\begin{center}

	{\Large \bf  {Celestial Superamplitudes} }  \\
	\vspace{0.3 cm}

	\vspace{25pt}

	{\mbox {\sf  \!\!\!\! Andreas~Brandhuber, Graham~R.~Brown, Joshua~Gowdy, Bill~Spence and 				Gabriele~Travaglini{\includegraphics[scale=0.05]{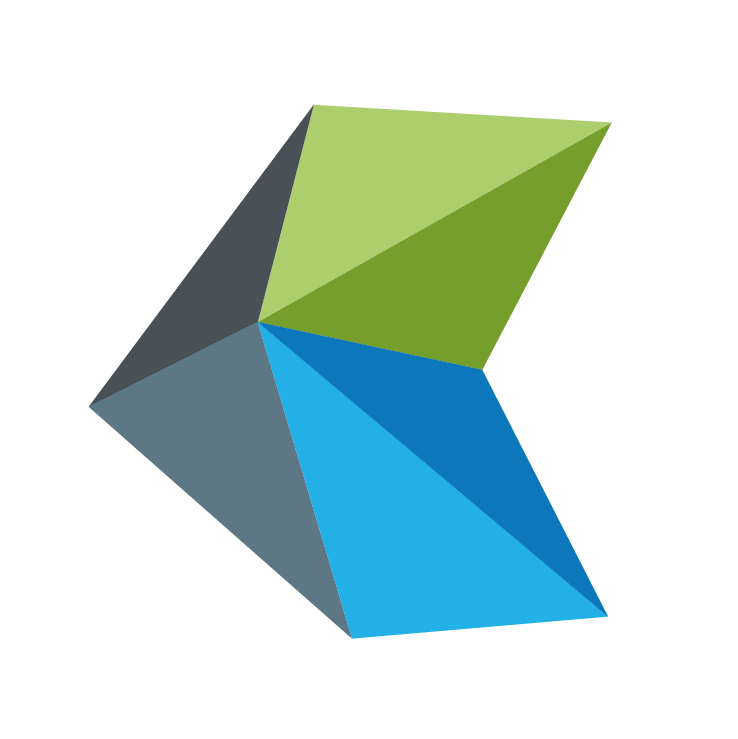}}
	}}
	\vspace{0.5cm}

	\begin{center}
		{\small \em
			Centre for Research in String Theory\\
			School of Physics and Astronomy\\
			Queen Mary University of London\\
			Mile End Road, London E1 4NS, United Kingdom
		}
		
	\end{center}

	\vspace{40pt}  

	{\bf Abstract}
\end{center}

\vspace{0.3cm}

\noindent

We study celestial amplitudes in (super) Yang-Mills theory using a parameterisation of the spinor helicity variables where their overall phase  is not fixed by the little group action. In this  approach the spin constraint $h-\bar{h}=J$ for celestial conformal primaries emerges naturally from a new Mellin transform, and the action of conformal transformations on celestial amplitudes is derived. Applying this approach to $\mathcal{N}\!=\!4$ super Yang-Mills, we show how the appropriate definition of on-shell superspace coordinates leads naturally to a formulation of chiral celestial superamplitudes and a representation of the generators of the  four-dimensional superconformal algebra on the celestial sphere, which by construction annihilate all tree-level celestial superamplitudes.
 
\vfill
\hrulefill
\newline
\vspace{-1cm}
${\includegraphics[scale=0.05]{Sagex.jpeg}}$~\!\!{\tt\footnotesize\{a.brandhuber, graham.brown, j.k.gowdy, w.j.spence, g.travaglini\}@qmul.ac.uk}

\setcounter{page}{0}
\thispagestyle{empty}
\newpage


\setcounter{tocdepth}{4}
\hrule height 0.75pt
\tableofcontents
\vspace{0.8cm}
\hrule height 0.75pt
\vspace{1cm}
\setcounter{tocdepth}{2}

\newpage
\section{Introduction}

The quest for secret symmetries of the $S$-matrix of gauge theory and gravity has been a continuous source of  surprises. An example is the discovery of the dual superconformal symmetry of $\cN\!=\!4$ supersymmetric Yang-Mills (SYM) theory, first conjectured in 
\cite{Drummond:2008vq} and subsequently proved at tree level in \cite{Brandhuber:2008pf}.
It emerged from earlier studies of iterative structures of MHV amplitudes in perturbation theory
\cite{Anastasiou:2003kj, Bern:2005iz, Drummond:2006rz} and at strong coupling \cite{Alday:2007hr}.
By replacing the standard kinematic variables with dual (super)momenta, a new duality of amplitudes with Wilson loops, valid at weak and strong coupling,
also became  apparent
\cite{Alday:2007hr,Drummond:2007aua,Brandhuber:2007yx,Drummond:2007cf}, with the dual symmetry acting as the usual conformal group on these new dual variables. 

Recently, the work of \cite{Pasterski:2016qvg,Pasterski:2017cbf,Pasterski:2017ylz}  suggested a new way to interpret scattering amplitudes of a generic four-dimensional theory
as correlators of a two-dimensional conformal field theory living at  null infinity of Minkowski spacetime, also known as the celestial sphere. This builds on an earlier, important  observation  that the Lorentz group \text{SO(3,1)}$\sim \sl2c$  of four-dimensional Minkowski space is mapped to the conformal group $\sl2c$ acting on the celestial sphere \cite{Terrell:1959zz,Penrose:1959vz,Held:1970kr} (see \cite{Oblak:2015qia} for a recent review). 
A fascinating consequence  of this viewpoint  is that the well-known soft theorems of gauge theories and gravity appear as Ward identities in the celestial conformal field theory  \cite{Lysov:2014csa,Cheung:2016iub, Donnay:2018neh,Himwich:2019dug,Fan:2019emx,Pate:2019mfs,Nandan:2019jas,Adamo:2019ipt,Puhm:2019zbl,Guevara:2019ypd,Fotopoulos:2020bqj}.
In fact, the natural group acting on the sphere at null infinity is the infinite-dimensional Bondi-van der Burg-Metzner-Sachs group \cite{Bondi:1962px,Sachs:1962zza}, which  is expected to be a symmetry of the dual celestial conformal field theory. 

The mapping from Minkowski space to the celestial sphere occurs through a Mellin transform which maps momentum eigenstates to boost eigenstates -- conformal wavepackets with well-defined conformal weights. The special features of this new basis have been discussed in \cite{Pasterski:2016qvg,Pasterski:2017cbf,Pasterski:2017ylz, Schreiber:2017jsr, Stieberger:2018edy,Arkani-Hamed:2020gyp}, in particular the boost eigenstates have unusual properties such as the non-decoupling of infrared and ultraviolet physics,  as emphasised in \cite{Arkani-Hamed:2020gyp}, which also gives a new angle on a possible  nonperturbative $S$-matrix  bootstrap by looking at the analytic properties of the celestial $S$-matrix. We also note  the recent works of \cite{Magnea:2021fvy,Gonzalez:2021dxw}, which study the universal infrared divergences  of celestial amplitudes in Yang-Mills theory, building on earlier work on gravity and quantum electrodynamics
\cite{Himwich:2020rro, Arkani-Hamed:2020gyp}. 
Most of the attention so far has been on  tree amplitudes, with the  exceptions of  \cite{Albayrak:2020saa,Gonzalez:2020tpi,Magnea:2021fvy,Gonzalez:2021dxw}, which discussed  the iterative structure of loop amplitudes or infrared divergences thereof. 

Work on supersymmetric theories has been mostly restricted  to minimal $\cN\!=\!1$ supersymmetric models
\cite{Fotopoulos:2020bqj, Pasterski:2021fjn}, 
whereas in this paper we focus on  the maximally supersymmetric $\cN\!=\!4$ theory,  its $S$-matrix and symmetries. 
One of the  questions we will answer is how to  systematically derive 
the action of the full superconformal group  on celestial (super)amplitudes, extending the work of \cite{Stieberger:2018onx} which presented the form of the generators of the four-dimensional conformal group on the celestial sphere. The Lorentz subgroup is of course implemented trivially as two-dimensional conformal transformations, however the remaining generators give rise to highly non-trivial constraints on a yet to be found celestial conformal field theory. The question  then naturally arises as to what are the celestial operators corresponding to the  generators of the full four-dimensional superconformal group. 

 In order to answer this question in a systematic way, in particular ensuring that the new generators in celestial space naturally obey the superconformal algebra, we find it useful to introduce a new ``chiral'' Mellin transform. In this approach we define the Mellin transform to celestial space in terms of  standard spinor-helicity variables  without modding out the little group redundancy occurring in the expression of a null momentum in terms of spinor variables. This has the advantage of making the translation between  Minkowski space generators and celestial generators immediate.  Using our new Mellin transform we define corresponding  celestial states and  (super)amplitudes (in the supersymmetric case). After performing the integration over the redundant little group phase, our celestial states and amplitudes are equal   to the conventional ones multiplied by a delta function imposing the spin constraint  $h-\bar{h} = J$ much in the same way as the momentum-conserving delta function in Minkowski spacetime. This allows us to give an unambiguous definition of the action of celestial operators and of their weights. In particular we discuss extensively how to compute weights of all the relevant operators in celestial space. We also present explicit expressions for tree-level MHV superamplitudes. Our results here agree with those of \cite{Kalyanapuram:2020aya}, although the specific forms  of the superamplitudes we present  are new. 

The rest of the paper is organised as follows. In Section~\ref{sec:2} we review the usual non-chiral Mellin transform and introduce our chiral celestial coordinates along with  their transformations under the Lorentz group. We also discuss the relation between these two sets of coordinates. 
In Section~\ref{sec:states and mellin transforms} we introduce chiral celestial states and amplitudes, and discuss the action of the Lorentz group on these quantities. 
Section~\ref{sec: CC algebra} is devoted to the derivation of the generators of the conformal group in Mellin space, showing the usefulness of our chiral representation. In this section we also explain how to determine the weights of operators in celestial space. 
After  briefly reviewing  basic properties of superamplitudes in $\cN\!=\!4$ SYM, we present in Section~\ref{sec:5} the formal derivation of celestial superamplitudes in this theory, together with explicit examples at three, four and five points, and  comments about the general $n$-point case.
Section  \ref{sec:Celestial SC algebra} is devoted to the derivation of explicit expressions for the full set of generators of the $\cN\!=\!4$ superconformal algebra.  
Finally, in Section~\ref{conclusions} we present our conclusions and outline possible lines of further research.
Appendix~\ref{app:A} expands on the derivation of weights of celestial operators, and in particular we present in Table \ref{tab:tableofweights} the list of weights of all relevant celestial operators. 

{\bf Note added:} after this work was completed  we became aware that Hongliang Jiang, also at Queen Mary University of London, had been working independently, and unknowingly to us due to Covid restrictions, on very similar topics \cite{Jiang:2021xzy}. Happily the results presented in our two papers are  in agreement although obtained in different ways. 

\section{Celestial coordinates}
\label{sec:2}
 \subsection{Review of non-chiral celestial coordinates}

As usual in four spacetime dimensions, we write massless momenta in terms of spinor helicity variables as 
$p_{\alpha \dot{\alpha}} = \ld{\alpha}\ltd{\alpha}$. 
 Little group transformations 
\begin{equation}
\label{LGtrans}
    \ld{\alpha}\rightarrow e^{i\phi} \ld{\alpha}, \qquad \tilde{\lambda}_{\dot{\alpha}}\rightarrow e^{-i\phi} \tilde{\lambda}_{\dot{\alpha}}\, , \end{equation}
leave a momentum invariant,  where we are restricting to the Minkowski slice where $p^{\mu}$ is real and $\ltd{\alpha} = \epsilon\, \ld{\alpha}^\ast$, where 
$\epsilon$ is an overall sign which is positive (negative) for outgoing (incoming) momenta. Out of the four real degrees of freedom in $\ld{\alpha}$, three are physical while its overall phase  $e^{i \theta}$ is  redundant. It can be set to zero by a little group transformation with $\phi=-\theta$.

Transitioning from the usual language of scattering amplitudes in four-dimensional Minkowski spacetime to conformal correlators on the celestial sphere $\cs2$ requires one to introduce new coordinates
\cite{Pasterski:2016qvg,Pasterski:2017cbf,Pasterski:2017ylz,Schreiber:2017jsr,Stieberger:2018edy,Stieberger:2018onx,Nandan:2019jas,Himwich:2019dug,Adamo:2019ipt}
\begin{align}
p_{\alpha \dot{\alpha}}:= \epsilon\, \omega \, q_{\alpha \dot{\alpha}}\, , 
    \end{align}
    where $q_{\alpha \dot{\alpha}}:= \xi_\alpha \tilde{\xi}_{\dot{\alpha}}$, with\footnote{We omit explicit spinor indices when there is no ambiguity.} 
\begin{align}
    \xi_\alpha
 \ = \ \begin{pmatrix}
			z \\
			1  
		\end{pmatrix}\, , \qquad 
	\tilde{\xi}_{\dot{\alpha}}\ = \ 	\begin{pmatrix}
			\zb \\
			1  
		\end{pmatrix}\ .
  \end{align}
  In vector notation 
  \begin{equation}\label{q-mu}
	 q^\mu\coloneqq\frac{1}{2}(1+ \abs{z}^2,-z-\zb,-i(z-\zb),1-\abs{z}^2)\, . 
\end{equation}
The  particle's energy is  $p^0=\frac{\omega}{2}(1+\abs{z}^2)$, however,  following standard convention  we will refer to $\omega$ as the \lq\lq energy\rq\rq. 

Following \cite{Arkani-Hamed:2020gyp}, one can parameterise  spinors in terms of the physical degrees of freedom:
\begin{equation}\label{eq:nonchiral spinors}
	\begin{split}
		& \lambda_\alpha= \epsilon\sqrt{\omega}\begin{pmatrix}
			z \\
			1  
		\end{pmatrix}
		 \, , 
		\quad\quad \,\,\,\,
		\tilde{\lambda}_{\dot{\alpha}}= \sqrt{\omega}\begin{pmatrix}
			\zb \\
			1  
		\end{pmatrix}
		\, , 
	\end{split}
\end{equation}
where
 a little group transformation was used to  set $\ld{2}=\sqrt{\omega}$ to be real. 
The spinors $\ld{\alpha}$ and $\ltd{\alpha}$ naturally transform in the (anti-)fundamental representation of $\sl2c$:
\begin{align}\label{lorentzlambda}
	\lambda\rightarrow \lambda^\prime = 
	\begin{pmatrix}
		a & b \\
		c & d  
	\end{pmatrix}\ \lambda\,   , \qquad 
	&\tilde{\lambda}\rightarrow \tilde{\lambda}^\prime 
	=\begin{pmatrix}
		\bar{a} & \bar{b} \\
		\bar{c} & \bar{d}  
	\end{pmatrix}\ \tilde{\lambda}\ . 
\end{align}
Hence, under a Lorentz transformation $p_{\alpha\dot\alpha}$ transforms as 
\begin{align}
\begin{split}
    p_{\alpha\dot{\alpha}} & \to 
\epsilon \, \omega \, (cz+d) (\bar{c}\bar{z}+\bar{d})
		\begin{pmatrix}
			\frac{az+b}{cz+d} \\
			1  
		\end{pmatrix}
		\begin{pmatrix}
			\frac{\bar{a}\bar{z}+\bar{b}}{\bar{c}\bar{z}+\bar{d}} \\
			1  
		\end{pmatrix}
		 \ = \epsilon \, \omega^\prime 
		\begin{pmatrix}
			z^\prime \\
			1  
		\end{pmatrix}
		\begin{pmatrix}
			\zb^\prime \\
			1  
		\end{pmatrix}
		\ = \epsilon \, \omega^\prime 
		\begin{pmatrix}
			 z^\prime\bar{z}^\prime&  z^\prime \\
			\bar{z}^\prime & 1  
		\end{pmatrix}\, , 
		\end{split}
\end{align}
where 
\begin{align}\label{sl2c}
	z':=\frac{a z +b}{cz+d}\, , \qquad \zb':=\frac{\bar{a}\zb+ \bar{b}}{\bar{c}\zb +\bar{d}}\, , \qquad ad-bc=1\, , 
\end{align}
which is a M\"obius transformation on $\cs2$, and   
\begin{equation}
\label{ometr}
	\omega \rightarrow \omega' = (cz+d)(\bar{c}\bar{z}+\bar{d})\omega \, .
\end{equation}
In the ($\omega,z, \bar{z})$ parameterisation we can then write Lorentz transformations on the spinors $\lambda_{\alpha}$ and $\tilde{\lambda}_{\dot\alpha}$ as \cite{Arkani-Hamed:2020gyp}
\begin{align}
\begin{split}
\label{eq: sl2Clambda}
	&\lambda\rightarrow \lambda^\prime = 
	\sqrt{\omega}\begin{pmatrix}
		a & b \\
		c & d  
	\end{pmatrix} \begin{pmatrix}
		z \\
		1  
	\end{pmatrix}= \left(\frac{\bar{c}\zb +\bar{d}}{cz+d}\right)^{-1/2}\sqrt{\omega'}\begin{pmatrix}
		z' \\
		1  
	\end{pmatrix},\\ 
	& \tilde{\lambda}\rightarrow \tilde{\lambda}^\prime 
	=\sqrt{\omega}\begin{pmatrix}
		\bar{a} & \bar{b} \\
		\bar{c} & \bar{d}  
	\end{pmatrix} \begin{pmatrix}
		\zb \\
		1  
	\end{pmatrix}=\left(\frac{\bar{c}\zb +\bar{d}}{cz+d}\right)^{1/2}\sqrt{\omega'}\begin{pmatrix}
		\zb' \\
		1  
	\end{pmatrix}\, .
	\end{split}
\end{align}
Note that just transforming the coordinates $\omega\rightarrow\omega'$ and $z\rightarrow z'$ is not  the same as performing a Lorentz transformation due to the additional phase factors in  \eqref{eq: sl2Clambda}, which appear as a consequence of the particular parameterisation chosen.
Finally we note the action of a  Lorentz transformation on $q$,
\begin{align}
	q^\mu\rightarrow \Lambda^\mu_{\,\,\nu}\, q^\nu=(cz+d)(\bar{c}\bar{z}+\bar{d})q^{\prime \mu}\ ,  
\end{align}
where $q_{\alpha \dot{\alpha}}^\prime= \xi_\alpha^\prime \tilde{\xi}_{\dot{\alpha}}^\prime$, with 
$    \xi^\prime_\alpha  
 \ = \ \begin{pmatrix}
			z^\prime \\
			1  
		\end{pmatrix}$ and  
$	\tilde{\xi}^\prime_{\dot{\alpha}}\ = \ 	\begin{pmatrix}
			\zb^\prime \\
			1  
		\end{pmatrix}$.
 %

\subsection{Chiral celestial coordinates}

In this paper  we  will employ  a different parameterisation of the spinors which does \textit{not} fix the little group redundancy. This has a number of advantages as we will see in the following. In this parameterisation, we define 
\begin{equation}\label{eq: chiral spinors}
    	\begin{split}
		& \lambda_\alpha:= \epsilon\,  u \begin{pmatrix}
			z \\
			1  
		\end{pmatrix}= \epsilon\, u\, \xi_\alpha\, 
		, 
		\qquad \,\,\,\,
		\tilde{\lambda}_{\dot{\alpha}}:= \ub\begin{pmatrix}
			\zb \\
			1  
		\end{pmatrix}
		= \ub \, \tilde{\celSp}_{\dot{\alpha}}\, , 
	\end{split}
\end{equation}
where  $u$ is a complex number such that 
\begin{align}
    u\ub = \omega
\ . 
\end{align}
It is immediate to relate this  ``chiral''  parameterisation of the spinors to the ``non-chiral'' one in \eqref{eq:nonchiral spinors}. Setting
\begin{equation}\label{eq: u in terms of omega}
    u= \sqrt{\omega}e^{i\theta}, \qquad \quad \bar{u}= \sqrt{\omega}e^{-i\theta}\, , 
\end{equation}
the connection between the two  is  simply 
\begin{align}
\label{newvsold}
    \lambda_{\rm C} = e^{i \theta}\lambda_{\rm NC}\ , \qquad \quad \tilde{\lambda}_{\rm C} = e^{-i \theta}\tilde{\lambda}_{\rm NC}\ , 
 \end{align}
with $e^{i \theta} = \sqrt{u/\bar{u}}$. Thus the parameterisation \eqref{eq: chiral spinors} contains  an additional degree of freedom. In the following we will regard  an amplitude as a function of the set of variables $\{z,\zb,u,\ub\}$. We will call these coordinates {\it chiral} celestial coordinates or just chiral coordinates, as opposed to the original {\it non-chiral} celestial coordinates $\{z,\zb,\omega\}$. Unlike in the non-chiral case, the chiral coordinates are a straightforward change of variables from the original spinor components $\ld{1},\ld{2},\ltd{1},\ltd{2}$. 
Furthermore, 
in the chiral parameterisation \eqref{eq: chiral spinors}, a Lorentz transformation of the spinors $\lambda_\alpha$ and $\tilde{\lambda}_{\dot\alpha}$ is written as  a standard $\sl2c$ transformation,
\begin{align}
\begin{split}
\label{chiralsl2Clambda}
	&\lambda\rightarrow \lambda^\prime = 
	u\begin{pmatrix}
		a & b \\
		c & d  
	\end{pmatrix} \begin{pmatrix}
		z \\
		1  
	\end{pmatrix}=  u' \begin{pmatrix}
		z' \\
		1  
	\end{pmatrix},\\ 
	& \tilde{\lambda}\rightarrow \tilde{\lambda}^\prime 
	=\ub\begin{pmatrix}
		\bar{a} & \bar{b} \\
		\bar{c} & \bar{d}  
	\end{pmatrix} \begin{pmatrix}
		\zb \\
		1  
	\end{pmatrix}=\ub'
	\begin{pmatrix}
		\zb' \\
		1  
	\end{pmatrix}\, , 
	\end{split}
\end{align}
where 
\begin{equation}\label{eq:M\"obius Transform}
    \begin{split}
        & u \rightarrow u'= (cz+d) u\, ,
        \qquad \quad \ub \rightarrow \ub' = (\cb \zb +\db) \ub\, .
    \end{split}
\end{equation}
It is instructive to contrast  \eqref{chiralsl2Clambda}  with \eqref{eq: sl2Clambda} -- in the former  the Lorentz transformation is equivalent to a M\"obius coordinate transformation, while in the latter an additional phase factor is needed.

\section{States and amplitudes in Mellin space}\label{sec:states and mellin transforms}
In momentum space, consider the scattering of asymptotic  states  labelled by their momentum $p^\mu$ and helicity $J$, 
where, under a  little group transformation \eqref{LGtrans},  
\begin{equation}\label{eq: old ket shift}
    \ket{p^\mu;J} \rightarrow \ket{p^\mu;J}'= \big(e^{i\phi}\big)^{-2J}\ket{p^\mu;J}.
\end{equation}
We can alternatively  label  states using spinor variables as $     |\ld{\alpha},\ltd{\alpha};J\rangle$. Then, in the chiral and non-chiral celestial coordinates we write 
\begin{subequations}
\begin{align}
    &\text{\bf Chiral:}   &&|z,\zb,u,\ub;J\rangle\, \coloneqq \, |
    u \, \xi_\alpha, 
    \ub \, \bar{\xi}_{\dot{\alpha}}; \, J\rangle\, , \label{eq: non-chiral state pre Mellin}\\
    &\text{\bf Non-chiral:}  &&|z,\zb,\omega;\, J\rangle\, \coloneqq\,  |
    \sqrt{\omega} \, \xi_\alpha, 
    \sqrt{\omega} \, \bar{\xi}_{\dot{\alpha}}; J\rangle\, . \label{eq: chiral state pre Mellin}
\end{align}
\end{subequations}
The two states are just related by the  little group transformation in  \eqref{newvsold},   
\begin{equation}\label{eq:c nc state relation pre mellin}
    \ket{z,\zb,u,\ub;J}= (e^{i\theta})^{-2J}\ket{z,\zb,\omega;J}\, , 
\end{equation}
with $e^{i \theta} = \sqrt{u/\bar{u}}$. 
This  extends immediately to a relation between amplitudes written in terms of chiral and non-chiral coordinates:
\begin{equation}\label{eq:c to nc amp pre mellin}
    A\big(\{z_i,\zb_i,u_i,\ub_i;J_i\}\big) = \prod_{j=1}^n\big(e^{i\theta_j}\big)^{-2J_j}A\left(\{z_i,\zb_i,\omega_i;J_i\}\right)\, , 
\end{equation}
 with  $e^{i \theta_j} = \sqrt{u_j/\bar{u}_j}$.
%
%
\subsection{Non-chiral  Mellin transform}\label{sec: nonchiral vs chiral mellin}
In the non-chiral approach of \cite{Pasterski:2016qvg,Pasterski:2017cbf,Pasterski:2017ylz}  one defines conformal primary states using a Mellin transform with respect to the energy of the particle
\begin{equation}
\label{MTfirst}
	\ket{ z,\zb,\Delta; J} \coloneqq \int_0^\infty\!d \omega \, \omega^{\Delta-1} \ket{ z,\zb,\omega; J}\, , 
\end{equation}
where $\Delta$ is the  two-dimensional conformal weight of the particle. Importantly, this state transforms as a conformal primary under the $\sl2c$ (or equivalently Lorentz) symmetry of the celestial sphere. From \eqref{eq: sl2Clambda} it follows that, under a Lorentz transformation, 
\begin{align}
\label{LTstate}
    \ket{ z,\zb,\omega; J}\to \ket{ z^\prime,\zb^\prime,\omega^\prime; J}\left(\frac{cz+d}{\bar{c}\bar{z} + \bar{d}}\right)^{-J}\, , 
\end{align}
hence, using \eqref{MTfirst} and \eqref{ometr}, 
\begin{equation}\label{boost Eigen Transform}
	\begin{split}
		\ket{ z,\zb,\Delta; J} &\rightarrow (cz+d)^{-\Delta -J} (\bar{c}\bar{z}+\bar{d})^{-\Delta+J}  \ket{ z',\zb',\Delta; J}\\
		& =\left(\frac{\partial z^{\prime}}{\partial z}\right)^{h} \left(\frac{\partial \zb^{\prime}}{\partial \zb}\right)^{\bar{h}}  \ket{z',\zb',\Delta;J}\, , 
	\end{split}
\end{equation}
which is the transformation law for a two-dimensional conformal primary with weights given in terms of $\Delta, J$ by 
\begin{equation}
	h =\frac{\Delta+J}{2}\, , \qquad \quad \bar{h}=\frac{\Delta-J}{2}\, . 
\end{equation}
Hence these conformal primary states can be labelled by $\{h,\bar{h}\}$ or $\{\Delta,J\}$ \cite{Pasterski:2016qvg}. 

\subsection{Chiral Mellin transform } \label{sec: Chiral Mellin Transform}
With the chiral variables $u, \bar u$ we now define a new  Mellin transform as
 \begin{equation}\label{eq: chiral Melin u}
    |z,\zb,h,\bar{h};J\rangle:=\frac{1}{2\pi i} \int_{\mathbb{C}}\!du\wedge d\ub \, u^{2h-1} \ub^{2\hb-1} |z,\zb,u,\ub;J\rangle\, .
\end{equation}
This is natural, since changing integration variables from $u,\bar{u}$ to $\omega,\theta$ using \eqref{eq: u in terms of omega},  and recalling \eqref{eq:c nc state relation pre mellin}, we find
\begin{equation}\label{eq:mellin change}
    \begin{split}
    |z,\zb,h,\hb;J\rangle
    &=\frac{1}{2\pi}\int_0^{2\pi}\!d\theta\  (e^{2i\theta})^{(h-\hb-J)}
    \int_0^\infty\!d\omega\, \omega^{h+\bar{h}-1} \ket{z,\zb,\omega;J}\, .  
    \end{split}
\end{equation}
Since we wish to consider conformal primaries with the usual half-integer helicity, we restrict $(h-\hb)\in \frac{1}{2}\mathbb{Z}$, and  we arrive at
\begin{equation}\label{eq:chiral mellin omega}
    |z,\zb,h,\hb;J\rangle=\delta_{h-\hb-J, 0}\int_0^\infty\!d\omega\, \omega^{h+\bar{h}-1}  |z,\zb,\omega;J\rangle\, . 
\end{equation}
 Thus the state vanishes when $h-\hb\neq J$, while when $h-\hb=J$ it  coincides with the non-chiral conformal primary state.

In the non-chiral case, a celestial amplitude is not initially written as a function of $h$ and $\hb$, since these quantities do not appear explicitly in the $\omega$ Mellin integral; in order to write it as such we must substitute according to the condition $h-\hb=J$ by hand. In contrast, in the chiral Mellin transform $h$ and $\hb$ are introduced a priori as free parameters varying within the constraint $(h-\hb) \in \frac{1}{2}\mathbb{Z}$;  the  chiral conformal primary is manifestly a function of $h,\hb$ which, however, is only non-vanishing when $h-\hb=J$.  The natural appearance of the spin constraint means that states have the expected weights $(h,\hb)=(\frac{\Delta+J}{2},\frac{\Delta-J}{2})$ selected by their helicity. The fact that $h$ and $\hb$ are now explicit in the chiral Mellin transform also makes the action of operators which shift conformal weight, such as  $e^{\partial_h/2}$,  unambiguous, as  discussed in  Section~\ref{sec: weights}.

Finally we note that under a Lorentz transformation 
\begin{align}
    \label{Linv-states}
     |z,\zb,u,\ub;J\rangle \to
     |z^\prime,\zb^\prime,u^\prime,\ub^\prime;J\rangle\, , 
     \end{align}
     which  follows simply from  \eqref{chiralsl2Clambda}.

\subsection{Celestial amplitudes: chiral vs non-chiral}
\label{sec:cvnc}
The discussion of chiral and non-chiral states extends naturally to amplitudes as follows: 
\begin{subequations}
\begin{align}
    &\text{\bf Chiral:} & &\tilde{A}^{\rm C}(\{z_i,\zb_i,h_i,\hb_i,J_i\})\coloneqq  \frac{1}{(2\pi i)^n}\prod_{j=1}^n \int_{\mathbb{C}} du_j\wedge d\ub_j \, u^{2h_j-1} \ub^{2\hb_j-1} A(\{z_i,\zb_i,u_i,\ub_i,J_i\})
    \label{chicc}\, ,\\
    &\text{\bf Non-chiral:} & &\widetilde{A}^{\rm NC}(\{z_i,\zb_i,\Delta_i,J_i\}) \coloneqq\prod_{j=1}^n \int_{0}^\infty d\omega_j \, \omega_j^{\Delta-1} A(\{z_i,\zb_i,\omega_i,J_i\})\, .
    \label{noncc}
\end{align}
\end{subequations}
We can compare these in a similar way to what was done for the states.
We change variables from $(u, \ub)$ to $(\omega, \theta)$ coordinates, and noting  \eqref{eq:c to nc amp pre mellin} we arrive at 
\begin{equation}\begin{split}
    \label{eq:c from nc amp post mellin}
    \widetilde{A}^{\rm C}(\{z_i,\zb_i,h_i,\hb_i,J_i\}) &= \prod_{j=1}^n \delta_{h_j-\hb_j -J_j,0} \int_{0}^\infty d\omega_j \, \omega_j^{h_j+\hb_j-1} A(\{z_i,\zb_i,\omega_i,J_i\})\\ & 
     = \prod_{j=1}^n \delta_{h_j-\hb_j -J_j,0}\, 
     \widetilde{A}^{\rm NC}(\{z_i,\zb_i,\Delta_i,J_i\})\, .
    \end{split}
\end{equation}
We conclude that the  non-chiral celestial amplitude is simply the chiral celestial amplitude with the Kronecker deltas $\delta_{h-\hb-J,0}$ stripped off for each particle. This is analogous to the situation where we  strip off the momentum conserving delta function from an amplitude and impose momentum conservation ``by-hand'' instead. For the rest of this paper we will work with chiral celestial amplitudes \eqref{chicc} and use the representation \eqref{noncc} when needed.%
\footnote{ From now on we will also drop the superscripts \text{C} and \text{NC} in the  celestial  amplitudes, unless we wish to make the distinction clear.} 

It  is also instructive to consider the 
 Lorentz transformation of a chiral conformal primary state \eqref{eq: chiral Melin u}. Recalling \eqref{Linv-states}, we have 
\begin{equation}\label{eq: chiral state Lorentz post mellin}
    \begin{split}
        |z,\zb,h,\hb;J\rangle & \rightarrow  \frac{1}{2\pi i} \int_{\mathbb{C}}\!du\wedge d\ub \, u^{2h-1} \ub^{2\hb-1} |z',\zb',u',\ub';J\rangle\\
        &  = (cz+d)^{-2h}(\cb \zb +\db)^{-2\hb}|z',\zb',h,\hb;J\rangle
        \, .
    \end{split}
\end{equation}
Thus the  state transforms under the two-dimensional conformal group with weights given by the parameters $h$ and $\hb$ appearing in the chiral Mellin transform. These considerations  extend to amplitudes --  
using the invariance of an  amplitude under Lorentz transformations and \eqref{eq: chiral state Lorentz post mellin} we have 
\begin{equation}\label{chiralcelestamplorentzinv}
    \begin{split}
        \widetilde{A}(\{z_i,\zb_i,h_i,\hb_i,J_i\})&= \prod_{j=1}^n (c z_j +d)^{-2h_j} (\bar{c}\zb_j +\bar{d})^{-2\hb_j} \widetilde{A}(\{z_i',\zb_i',h_i,\hb_i,J_i\})
        \, , 
    \end{split}
\end{equation}
which is just the transformation  for a correlator of $n$ two-dimensional conformal primaries, as expected. 

\section{From chiral coordinates to the celestial conformal algebra}\label{sec: CC algebra}
As a first demonstration of the use of chiral coordinates and the  chiral Mellin transform, we  rederive the celestial representation of the four-dimensional conformal generators first presented in \cite{Stieberger:2018onx}.

\subsection{Celestial conformal generators}

Our first  goal is to find representations of $\ld{\alpha}$ and  $\ltd{\alpha}$  on the celestial sphere, that is derive  their induced action on celestial amplitudes in Mellin space. In chiral coordinates this is straightforward. Firstly, we change variables from $\{\ld{\alpha},\ltd{\alpha}\}$ to $\{z,\zb,u,\ub\}$, with 
\begin{equation}
\begin{aligned}
        &\ld{\alpha}=u\begin{pmatrix}
            z\\
            1
        \end{pmatrix}, &\quad 
        &\ltd{\alpha}= \ub \begin{pmatrix}
            \zb\\
            1
        \end{pmatrix}\, , \\
        & \pld{\alpha}\coloneqq\partdiv{}{\lu{\alpha}} =\frac{1}{u}\begin{pmatrix}
            u\partial_u-z\partial\\
            -\partial
        \end{pmatrix}\, , &\quad \qquad  &\pltd{\alpha}\coloneqq\partdiv{}{\ltu{\alpha}} = \frac{1}{\ub}\begin{pmatrix}
            \ub\partial_{\ub}-\zb\bar{\partial}\\
            -\bar{\partial}
        \end{pmatrix},
\end{aligned}
\end{equation}
where  
$\partial_{u}=\partial / \partial u$ 
and 
$\partial_{\bar{u}}=\partial / \partial \bar{u}$. 
 Next we define the action of these objects on celestial amplitudes using the chiral Mellin transform. For instance, for  $\ld{\alpha}$ 
\begin{equation}
        \ld{\alpha}\widetilde{A}= \frac{1}{2\pi i} \int_{\mathbb{C}}\!du \wedge d\ub \, u^{2h-1} \ub^{2\hb-1} u\begin{pmatrix}
            z\\
            1
        \end{pmatrix} A
                            = \begin{pmatrix}
            z\\
            1
        \end{pmatrix} \frac{1}{2\pi i} \int_{\mathbb{C}}\!du \wedge d\ub \, u^{2(h+1/2)-1} \ub^{2\hb-1} A\, .
\end{equation}
We conclude that, when transformed to $\cs2$,  $\ld{\alpha}$ is represented by the operator%
\footnote{We use the same symbols for the operator representations of objects in both momentum space and celestial space; the difference will be clear from the context.}
$    \ld{\alpha}=   \begin{pmatrix}
            z\\
            1
        \end{pmatrix} e^{\frac{1}{2}\partial_{h}}$,
where the exponential operator shifts $h \rightarrow h+\frac{1}{2}$. 
Thus we find
\begin{equation}\label{celestspinorgens}
    \begin{aligned}
            &\ld{\alpha}=\begin{pmatrix} z\\1 \end{pmatrix} e^{\frac{1}{2}\partial_h}\, , &\quad &\ltd{\alpha}=\begin{pmatrix} \zb\\1 \end{pmatrix}
            e^{\frac{1}{2}\partial_{\hb}}\, , \\
            &\pld{\alpha}=-\begin{pmatrix}
            2h-1+z\partial\\ \partial
            \end{pmatrix} e^{-\frac{1}{2} \partial_h}\, , &\quad
            &\pltd{\alpha}=-\begin{pmatrix}
            2\hb-1+\zb\partialb\\ \partialb
    \end{pmatrix} e^{-\frac{1}{2} \partial_{\hb}}\, .
    \end{aligned}
\end{equation}
We can then immediately write down the generators of the conformal algebra from their momentum space counterparts using the replacements \eqref{celestspinorgens}. 

These are:
\begin{align}
    p_{\alpha\dot{\alpha}}&=\ld{\alpha}\ltd{\alpha}  =\begin{pmatrix}
		z\zb & z \\
		\zb & 1 
	\end{pmatrix}_{\alpha\dot{\alpha}} e^{\frac{1}{2}(\partial_h+\partial_{\hb})}\, , 
\\
    m_{\alpha\beta}&=\ld{(\alpha}\pld{\beta)}=\begin{pmatrix}
         -2zh -z^2\partial & -h-z\partial\\ \cr
         -h-z\partial & -\partial
    \end{pmatrix}_{\alpha\beta}, 
    \\
    \bar{m}_{\dot{\alpha}\dot{\beta}}&=\tilde{\lambda}_{(\dot{\alpha}}\tilde{\partial}_{\dot{\beta})}=\begin{pmatrix}
         -2\zb\hb -\zb^2\partialb & -\hb-\zb\partialb\\ \cr
         -\hb-\zb\partialb & -\partialb
    \end{pmatrix}_{\dot{\alpha}\dot{\beta}}, 
\\
 k_{\alpha\dot{\alpha}}&=\pld{\alpha}\pltd{\alpha}=\begin{pmatrix}
    (2h-1+z\partial)(2\hb-1+\bar{z}\bar{\partial}) & (2h-1+z\partial)\bar{\partial}\\ \cr
(2\hb-1+\bar{z}\bar{\partial})\partial & \partial\bar{\partial}
\end{pmatrix}_{\alpha\dot{\alpha}} e^{-\frac{1}{2}(\partial_{h}+\partial_{\hb})}\, , 
\\
    d&=\frac{1}{2}\lu{\alpha}\pld{\alpha}+ \frac{1}{2}\ltu{\alpha}\pltd{\alpha}+1=-(h+\hb-1)\,  .
\end{align}
These  match the generators given in \cite{Stieberger:2018onx}. Note that they  automatically satisfy the conformal algebra%
 \footnote{The superconformal algebra of $\cN=4$ SYM can be found in e.g. \cite{Drummond:2008vq,Drummond:2009YS}. }. 
 This follows from the fact that the celestial space operators representing the spinors and the spinor derivatives have the same basic commutation relations as the original momentum space spinors and their derivatives.
 
The last property we wish to check is that the celestial conformal generators  annihilate conformally invariant amplitudes, e.g. tree amplitudes in pure Yang-Mills  theory. Once again  this is  immediate:  all the celestial generators above have the property that we can push them back through the Mellin transform to obtain the original momentum space versions. 
In summary, the celestial generators satisfy all of the properties one would expect: they close the algebra and annihilate amplitudes. 

We comment that deriving these generators using non-chiral coordinates is not as straightforward as above,  as it involves relating four variables,  $\{\ld{\alpha},\ltd{\alpha}\}$ to three, $\{z,\zb,\omega\}$. To fix this in non-chiral coordinates we must require
\begin{equation}
    \ld{\alpha}= \sqrt{\omega}\begin{pmatrix}
        z\\
        1
    \end{pmatrix}, 
    \qquad \ltd{\alpha}= \sqrt{\omega}\begin{pmatrix}
        \zb\\
        1
    \end{pmatrix}\quad \implies \ld{2}=\ltd{2}\, .
\end{equation}
Due to this constraint,   the spinors $\ld{\alpha}$ and $\ltd{\alpha}$, as well as their   derivatives $\pld{\alpha}$, $\pltd{\alpha}$, can  no longer be treated as independent, which considerably complicates the analysis.

\subsection{Weights of celestial operators}\label{sec: weights}
In using  chiral coordinates we no longer have explicit weight factors appearing when we perform a Lorentz transformation in momentum space 
(see  \eqref{chiralsl2Clambda}, as opposed to its non-chiral counterpart  \eqref{eq: sl2Clambda}).  Celestial amplitudes and the spinors they are built from will of course still  have weights under the two-dimensional conformal group on the celestial sphere. By relating the chiral Mellin transform (which treats $h$ and $\hb$ as free parameters) to the non-chiral Mellin transform, the weights $h,\hb$ of a celestial amplitude were found in Section \ref{sec: nonchiral vs chiral mellin} to be constrained to be $(h,\hb) = (\frac{\Delta+J}{2}, \frac{\Delta-J}{2})$. We now show how this mechanism also assigns weights to the celestial conformal generators according to their helicity.

Consider a celestial spinor acting on a state $|z,\zb,h,\hb;\, J\rangle$, 
\begin{align}\label{spinoractingonamp-old}
    \begin{split}
    &\ld{\alpha}|z,\zb,h,\hb;J\rangle = \frac{1}{2\pi i} \int_{\mathbb{C}}\!du\wedge d\ub\ u^{2h-1}\ub^{2\hb-1}\; u\begin{pmatrix}
        z\\1
    \end{pmatrix}_{\alpha} \, |z,\zb,u,\ub;J\rangle \ . 
    \end{split}
    \end{align}
    Under a Lorentz transformation, spinors and states transform as in 
\eqref{chiralsl2Clambda} and \eqref{Linv-states},  respectively, hence under a Lorentz transformation we have
\begin{align}\label{spinoractingonamp}
    \begin{split}
    \ld{\alpha}|z,\zb,h,\hb;J\rangle 
    &\longrightarrow 
    \frac{1}{2\pi i}\int_{\mathbb{C}}\!du\wedge d\ub\  u^{2h-1}\ub^{2\hb-1}\, u^\prime \begin{pmatrix}
        z'\\1
    \end{pmatrix}_{\alpha} |z',\zb',u',\ub';J\rangle\\
& =     (cz+d)^{-2h}(\cb\zb+\db)^{-2\hb}\, \frac{1}{2\pi i}\int_{\mathbb{C}}\!du'\wedge d\ub'\  u^{\prime \;2h}\ub^{\prime\;2\hb-1} \begin{pmatrix}
        z'\\1
    \end{pmatrix}_{\alpha} |z',\zb',u',\ub';J\rangle\\
    &= (cz+d)^{-2h}(\cb\zb+\db)^{-2\hb}\  \ld{\alpha}'\ |z',\zb',h,\hb;J\rangle \, ,
    \end{split}
\end{align}
where in the last line $\lambda_\alpha^\prime = \begin{pmatrix}
        z'\\1
    \end{pmatrix}_{\alpha}e^{\frac{1}{2}\partial_h} $. 
From \eqref{spinoractingonamp} we see that the celestial spinor operator acting on a state, $ \ld{\alpha}\ |z,\zb,u,\ub;J\rangle$, has seemingly the same weight factors as $|z,\zb,u,\ub;J\rangle$ and this is also the case for all of the chiral operators in \eqref{celestspinorgens}. This is an incorrect conclusion.  
Indeed, this argument misses the fact that the parameters $h,\hb$ must acquire values that satisfy the spin constraint arising from the  Kronecker delta. As such, to ascertain the weights of $\ld{\alpha}\ |z,\zb,u,\ub;J\rangle$ we must also consider the action of the operator $\ld{\alpha}$ on the spin constraint.
We now consider the action of a spinor on a state again,  making the connection to the non-chiral Mellin transform explicit,
\begin{align}
\label{412}
    \begin{split}
    \ld{\alpha}|z,\zb,h,\hb;J\rangle 
    &=\begin{pmatrix}
        z\\1
    \end{pmatrix}_{\alpha} e^{\frac{1}{2}\partial_h}\ \delta_{h-\hb-J,0}\ |z,\zb,\Delta;J\rangle 
    \\
    &=\begin{pmatrix}
        z\\1
    \end{pmatrix}_{\alpha}\delta_{h-\hb-\left(J-\frac{1}{2}\right),0}\int_0^{\infty} d\omega  \ \omega^{h+\hb+\frac{1}{2}-1} \, |z,\zb,\omega;J\rangle
   \\ &=\begin{pmatrix}
        z\\1
    \end{pmatrix}_{\alpha}\delta_{h-\hb-\left(J-\frac{1}{2}\right),0}\  |z,\zb,\Delta + \frac{1}{2};J\rangle\, .
    \end{split}
\end{align}
From \eqref{412} we can read off 
\begin{align}
    \begin{split}
    &h-\hb=J-\frac{1}{2}\, , \qquad \quad h+\hb+\frac{1}{2}=\Delta+\frac{1}{2}
    \, ,
    \end{split}
    \end{align}
therefore
    \begin{align}
        \begin{split}
     h=\frac{\Delta+J}{2}-\frac{1}{4}\, , \qquad \quad \hb=\frac{\Delta-J}{2}+\frac{1}{4}\, .
    \end{split}
\end{align}
These values can now be plugged in \eqref{spinoractingonamp} to find the weights of the operator $\lambda_\alpha$ in celestial space. From this we see that $\ld{\alpha}$ has shifted the weights by $(-\frac{1}{4},\frac{1}{4})$, as expected  from  an object with helicity $-\frac{1}{2}$. 
This  carries over to the other celestial generators, and we provide further examples in Appendix \ref{app:A} along with a complete table of weights, Table \ref{tab:tableofweights}. Note that generators such as $p_{\alpha\dot\alpha}$ and $k_{\alpha\dot\alpha}$ are weightless since they do not feature operators that act on the spin constraint $h-\hb=J$.


\section{From on-shell superspace to celestial superamplitudes}\label{sec:5}

\subsection{\texorpdfstring{Lightning review of $\mathcal{N}=4$}{N=4} superamplitudes }\label{sec: SYM review}

We begin with a short review of the amplitudes in $\cN=4$ SYM. In this theory one can  package all  amplitudes with fixed total helicity and  number of particles $n$ into a superamplitude. This depends on some   auxiliary Gra{\ss}mann variables $\eta_i^A$, one for each particle (labelled by $i$), with $A=1,\ldots , 4$ being an index of the fundamental representation of the $\SU4$ $R$-symmetry group.  By expanding  a superamplitude in these   fermionic variables, one can obtain the various component amplitudes. The precise correspondence  can be derived  from the Nair super-creation operator (often called the Nair superwavefunction) \cite{Nair:1988bq} which incorporates the creation operators of the physical states,
\begin{equation}\label{eq:super-field def}
    \Phi(\lambda,\tilde{\lambda},\eta) = G_{+} + \eta^A \Gamma_{A} + \frac{1}{2!}\eta^A\eta^B S_{AB} + \frac{1}{3!} \varepsilon_{ABCD}\eta^B\eta^C\eta^D \bar{\Gamma}^A + \frac{1}{4!}\varepsilon_{ABCD}\eta^A\eta^B\eta^C\eta^D G_{-}\, .
\end{equation}
Then, in order to select a state with  helicity $J_i$, one needs to pick the term with $2-2J_i$ powers of $\eta_i$ in the Taylor expansion of the superamplitude. 
Note that $\Phi(p,\eta)$ has uniform helicity $+1$ since we assign the $\eta^A$ variables helicity $+1/2$. 

$\cN\!=\!4$ SYM is a    superconformal theory, and we now briefly review the generators of this symmetry.  Firstly,  there are the  supersymmetry generators $q_{\alpha}^A$, $\bar{q}_{\dot{\alpha}A}$, which satisfy 
\begin{equation}\label{eq:supersymmerty defining comutator}
    \{q_{\alpha}^A,\bar{q}_{\dot{\alpha} B}\}\, = \,  \delta_B^A\,  \lambda_{\alpha}\tilde{\lambda}_{\dot{\alpha}}\, .
\end{equation}
Importantly, on-shell superspace allows us to realise these generators as 
\begin{equation}\label{supermomentum}
    q_\alpha^A= \ld{\alpha}\eta^A, \qquad \quad \bar{q}_{\dot{\alpha}A}= \ltd{\alpha}\partial_A. 
\end{equation}
In addition we have the generators of dilations and special conformal transformations  \cite{Witten:2003nn}
\begin{equation}
    d =\frac{1}{2}\lu{\alpha}\partial_{\alpha} + \frac{1}{2}\ltu{\alpha}\partial_{\alpha} + 1, \qquad \quad k_{\alpha\dot{\alpha}} = \partial_{\alpha}\partial_{\dot{\alpha}}\, , 
\end{equation}
together with two conformal supersymmetry generators
\begin{equation}
    s_{\alpha A} = \partial_\alpha \partial_{A}, \qquad\quad \bar{s}_{\dot{\alpha}}^A = \partial_{\dot{\alpha}} \eta^A.
\end{equation}
Since the helicity of each supermultiplet created by $\Phi(\lambda,\tilde{\lambda},\eta)$ is equal to one, the operator 
\begin{equation}
    \cJ = \frac{1}{2}\big( -\lu{\alpha} \partial_{\alpha} + \tilde{\lambda}^{\dot{\alpha}}\partial_{\dot{\alpha}} + \eta^A\partial_{A} \big)\, , 
\end{equation}
is such that  for each particle $i$ one has \cite{Witten:2003nn}  
\begin{equation}
\label{superhel}
     \cJ_i \cA_n(\{\ld{j}, \tilde{\lambda}_{j}, \eta_j\}) = \cA_n(\{\ld{j}, \tilde{\lambda}_{j}, \eta_j\})\, .
\end{equation}
The algebra also has a  central charge operator $c= 1-\cJ$ which for each particle  annihilates the superamplitude: 
\begin{equation}
   c_i \cA_n(\{\ld{j}, \tilde{\lambda}_{j}, \eta_j\}) = 0\, .
\end{equation}
Finally the  $\SU4$  generators are
\begin{equation}
     r^A_{\,\,B}=\eta^A \partial_B \, - \, \frac{1}{4} \, \delta_B^A \, \eta^C\,  \partial_C.
\end{equation}
 Our next task is  to find expressions for these generators on  the celestial sphere.

\subsection{Celestial on-shell superspace}
To consider celestial superamplitudes we must first extend the celestial sphere to include a  Gra{\ss}mann variable and hence define a $\cs2$ superspace. This was done for the $\mathcal{N}=1$ case in \cite{Fotopoulos:2020bqj}, we now do this for  $\mathcal{N}=4$ SYM  using chiral coordinates in the on-shell superspace formalism reviewed in  Section \ref{sec: SYM review}. 

To begin with,  we  consider a Mellin transform of the superwavefunction \eqref{eq:super-field def}, as we did in Section \ref{sec: Chiral Mellin Transform} for single-particle states. However, before we do this we must consider the transformation properties of $\eta^A$, which is invariant under a Lorentz transformation, and  transforms under a little group rotation of the corresponding particle since it carries helicity $+1/2$. Thus, in a similar manner to the positive-helicity spinor $\ltd{\alpha}$, we  replace  $\eta^A$ in terms of a new Gra{\ss}mann variable $\tau^A$ which is invariant under the little group together with a phase that carries the little group transformation: 
\begin{equation} \label{eq: eta to upeta relation}
    \eta^A\to  e^{-i\theta}\tau^A= \sqrt{\frac{\ub}{u}} \tau^A\, .
\end{equation}
Using the Lorentz invariance of $\eta^A$, the definition   \eqref{eq: eta to upeta relation} and  \eqref{eq:M\"obius Transform}, we see that $\tau^A$ must change with the following M\"obius  transformation under a Lorentz transformation:
\begin{align}
\label{boh}
    \tau^{\prime A}=\left(\frac{cz+d}{\cb\zb+\db}\right)^{\frac{1}{2}}\tau^A\, , 
\end{align}
 while it is by construction invariant under the little group. 
Note that \eqref{boh} parallels \eqref{eq: sl2Clambda}.

From here we define the following  conformal primary celestial superwavefunction:  
\begin{equation}
\begin{split}
    &\tilde{\Phi}(z,\zb, h ,\hb, \tau):= \frac{1}{2\pi i} \int_{\mathbb{C}}\!du\wedge d\ub \, u^{2h-1} \ub^{2\hb-1}(\ket{z,\zb,u,\ub,+1} +\eta^A \ket{z,\zb,u,\ub, +1/2}_A + \cdots)\, ,
\end{split}
\end{equation}
where on the right-hand side of this equation $\eta$ should be regarded as the function of $\tau$ obtained by performing the replacement in  \eqref{eq: eta to upeta relation}.  
The reason behind this definition becomes clear once we relate this to the non-chiral Mellin transform. Indeed,  doing so we see that the superwavefunction (and hence the superamplitudes) has a uniform helicity weight $+1$: 
\begin{equation}
    \tilde{\Phi}(z,\zb, h ,\hb, \tau)=  \delta_{h-\hb-1,0} \int_0^{\infty}\!d\omega \, \omega^{h+\hb-1}(\ket{z,\zb,\omega,+1} +\tau^A \ket{z,\zb,\omega, +1/2}_A + \cdots)\,, 
\end{equation}
where we have used \eqref{eq:c nc state relation pre mellin}.
Note that had we not performed the replacement \eqref{eq: eta to upeta relation}  
then each term in the superwavefunction would have different weights, 
and  we could not have interpreted it as a conformal primary. 

As with the spinors and their derivatives, we can write $\eta_A$ and $\partial_A$ as operators on the celestial sphere by pulling them through the Mellin transform
\begin{equation}\label{eq: etagens}
    \eta^A = \tau^A e^{(-\partial_h+\partial_{\hb})/4}, \qquad \quad \partial_A= \check{\partial}_A e^{(\partial_h-\partial_{\hb})/4},
\end{equation}
where we define $\check{\partial}_A = \partdiv{}{\tau^A}$.  Using these operators we can extract the  conformal primaries \ for all particles  from the celestial superwavefunction by simply taking fermionic derivatives. For example, for the gluino we have
\begin{equation}
    \begin{split}
    |z,\zb,h,\hb,+1/2\rangle_A&= \partial_A \tilde{\Phi}(z,\zb, h ,\hb, \tau)\lvert_{\tau=0}
    \, = \, \check{\partial}_A e^{(\partial_h-\partial_{\hb})/4}\tilde{\Phi}(z,\zb, h ,\hb, \tau)\lvert_{\tau=0}\\
    &=\delta_{h-\hb-\frac{1}{2},0}\int_0^{\infty}\!d\omega \, \omega^{h+\hb-1}\ket{z,\zb,\omega, +1/2}_A.
    \end{split}
\end{equation}

\subsection{Celestial superamplitudes}
The chiral Mellin transform of a superamplitude proceeds as a natural extension of the non-supersymmetric case. A celestial superamplitude is given by
\begin{align}
    \widetilde{\cA}_n(\{z_i,\zb_i,h_i,\hb_i,\tau_i\}):=\frac{1}{(2\pi i)^n}\prod_{i=1}^{n} \int_{\mathbb{C}}\,  du_i\wedge d\ub_i \  u_i^{2h_i-1}\ub_i^{2\hb_i-1} \cA_n \left(\left\{u_i,\ub_i,z_i,\zb_i,\sqrt{\frac{\ub_i}{u_i}}\tau_i \right\}\right)\, .
\end{align}
We can again relate this to the non-chiral Mellin transform in the $\omega$ coordinate similarly to what was done in Section \ref{sec:cvnc}. The result is
\begin{equation}\label{eq: SAmp c to nc}
    \widetilde{\cA}_n(\{z_i,\zb_i,h_i,\hb_i,\tau_i\})
    =\prod_{i=1}^n \delta_{h_i-\hb_i-1,0}\int_0^{\infty}\!d\omega_i\,  \omega_i^{h_i+\hb_i-1}  \cA_n \left(\left\{\omega_i,z_i,\zb_i,\tau_i \right\}\right),
\end{equation}
where we have used \eqref{superhel}.
Note the appearance of Kronecker deltas enforcing the spin constraint $h_i-\hb_i=1$ for every particle $i$. \eqref{eq: SAmp c to nc} is the supersymmetric generalisation of \eqref{eq:c from nc amp post mellin}. 
On  the support of the Kronecker delta, the celestial superamplitude transforms as a two-dimensional conformal correlator with weights now given by $(h_i,\hb_i)= (\frac{\Delta_i+1}{2}, \frac{\Delta_i-1}{2})$. Under Lorentz transformations we have
\begin{align}\label{eq:scampTransform}
\begin{split}
		&\widetilde{\mathcal{A}}_n(\{z_i,\zb_i,h_i,\hb_i,\tau_i\})\\
	&	=\prod_{i=1}^n \left(c z_i +d\right)^{-\Delta_i-1} \left(\bar{c}\zb_i +\bar{d}\right)^{-\Delta_i+1}  \widetilde{\mathcal{A}}_n\bigg(\bigg\{\frac{a z_i +b}{c z_i + d},\frac{\bar{a}\zb+ \bar{b}}{\bar{c}\zb +\bar{d}},h_i,\hb_i,\left(\frac{\cb\zb+\db}{cz+d}\right)^{-\frac{1}{2}}\tau_i\bigg\}\bigg)\, .
	\end{split}
\end{align}
We can examine the infinitesimal version of  \eqref{eq:scampTransform} to obtain the generators of the Lorentz symmetry on celestial superamplitudes. In order to do so we first note that 
a M\"{o}bius transformation \eqref{sl2c} is now accompanied by the  transformation   
$\tau\to \tau^{\prime A}$ defined in 
\eqref{boh}.
The infinitesimal transformation has then the form 
\begin{align}
    z'= z +\epsilon(z)\, , \qquad \zb' = \zb + \bar{\epsilon}(\bar{z})\, , \qquad \tau^\prime= \tau + \kappa (z, \bar{z}) \tau\, , 
\end{align}
with $\kappa = \frac{1}{4}\bar\partial\bar\epsilon - \frac{1}{4}\partial\epsilon$.
The corresponding generator acting on each particle is then (omitting the particle label) 
\begin{equation}
\label{520}
	\begin{split}
	&	h \partial\epsilon \, + \, \epsilon\partial\,  + \, \bar{h} \bar{\partial}\bar{\epsilon} \, + \, \bar{\epsilon}\bar{\partial} + \kappa \, \tau^A\check{\partial}_A
	\\  = \ &
		\Big(h-\frac{1}{4} \tau^A\check{\partial}_A\Big)
		\partial\epsilon + \epsilon\partial + 
		\Big(\bar{h} +\frac{1}{4} \tau^A\check{\partial}_A\Big)\bar{\partial}\bar{\epsilon} + \bar{\epsilon}\bar{\partial} \, .
	\end{split}
\end{equation}
Thus we obtain the celestial superspace  generators from those defined earlier in celestial space by simply shifting $h$ and $\bar h$ as 
\begin{align}
\label{hshift}
   h\to h-\frac{1}{4}\tau^A\check{\partial}_A\, , \qquad  \quad \bar{h} \to \hb+\frac{1}{4}\tau^A\check{\partial}_A\, .
\end{align}
As an example, we  have the following  Lorentz generators acting on superamplitudes:
\begin{align}\label{eq: super Lorentz}
\begin{split}
    m_{\alpha\beta}=\begin{pmatrix}
         -2z(h-\frac{1}{4}\tau^A\check{\partial}_A) -z^2\partial & -(h-\frac{1}{4}\tau^A\check{\partial}_A)-z\partial\\ \cr
         -(h-\frac{1}{4}\tau^A\check{\partial}_A)-z\partial & -\partial
    \end{pmatrix}_{\alpha\beta},\\ \cr
    \bar{m}_{\dot{\alpha}\dot{\beta}}=\begin{pmatrix}
         -2\zb(\hb+\frac{1}{4}\tau^A\check{\partial}_A) -\zb^2\partialb & -(\hb+\frac{1}{4}\tau^A\check{\partial}_A)-\zb\partialb\\ \cr
         -(\hb+\frac{1}{4}\tau^A\check{\partial}_A)-\zb\partialb & -\partialb
    \end{pmatrix}_{\dot{\alpha}\dot{\beta}}.
    \end{split}
\end{align}
These generators feature the weights of the superamplitude $(h,\hb)= (\frac{\Delta+1}{2},\frac{\Delta-1}{2})$ but shifted in such a way as to subtract off the weights of  $\tau^A$ (or equivalently $\eta^A$), which, since it carries helicity $+\frac{1}{2}$, are just $(\frac{1}{4}, -\frac{1}{4})$. In these generators, the 
weight-counting operator acts on each term in the superamplitude and measures the weights of \textit{only the component amplitude} in that term. Indeed this must be the case since the Lorentz generators act on the spinors alone and annihilate the superamplitude precisely by annihilating each component amplitude.

Note that the combination $\tau^A\check{\partial}_A$ is little group invariant and therefore we can replace it with  $\eta^A\partial_A$. We leave this expression as it is to highlight the role $\tau^A$ plays in the chiral coordinates and also in the celestial superconformal algebra in Section \ref{sec:Celestial SC algebra}.

\subsection{Examples of celestial superamplitudes}
Here we derive the supersymmetric three-, four- and five-point 
tree-level MHV superamplitudes using the Mellin transform.\footnote{This section on explicit MHV celestial superamplitudes partly overlaps with work carried out recently in \cite{Kalyanapuram:2020aya} although the formul\ae\ at four- and five-points \eqref{bills4point}, \eqref{bills5point}, are new.}  We also make some remarks on the general case.  We will work with the chiral Mellin transform using its relationship with the standard non-chiral transform \eqref{eq:c from nc amp post mellin}.
To begin with we write the three-point MHV tree-level superamplitude in terms of non-chiral celestial coordinates,
\begin{equation}
\begin{split}
    \cA_3^{\text{MHV}}&= \delta^{(4)}(p_1+p_2-p_3) \frac{\delta^{(8)}(q_{1\alpha}^A+q_{2\alpha}^A-q_{3\alpha}^A)}{\agl{1}{2}\agl{2}{3}\agl{3}{1}}\\
   & =\delta^{(4)}(p_1+p_2-p_3) \frac{\delta^{(8)}(\sqrt{\omega_1} \xi_1 \tau_1^A +\sqrt{\omega_2} \xi_2\tau_2^A -\sqrt{\omega_3}\xi_3\tau_3^A)}{-\omega_1\omega_2\omega_3 z_{12}z_{23}z_{31}}\, ,
\end{split}
\end{equation}
where we take particles 1 and 2 incoming and 3 outgoing. To evaluate the Mellin transform with respect to each $\omega_i$ we first rewrite the momentum conserving delta function as follows,
\begin{equation}
    \begin{split}
        \delta^{(4)}(p_1+p_2-p_3)&= \delta^{(4)}(\omega_1\xi_1\tilde{\xi}_1 +\omega_2\xi_2\tilde{\xi}_2-\omega_3\xi_3\tilde{\xi}_3)\\
        &= \frac{4}{\omega_3^2 z_{23}z_{31}}\delta(\zb_{31})\delta(\zb_{23})\delta\Big(\omega_1-\omega_3\frac{z_{23}}{z_{21}}\Big) \delta\Big(\omega_2-\omega_3\frac{z_{31}}{z_{21}}\Big)\, .
    \end{split}
\end{equation}
The first two delta functions here imply the constraint $\tilde{\lambda}_1\sim \tilde{\lambda}_2\sim \tilde{\lambda}_3$ as expected for three-point MHV amplitudes\footnote{Technically this means we must work with complexified momenta with $z$ and $\zb$ independent at three points.}. The remaining two delta functions trivialise the $\omega_1$ and $\omega_2$ integrals to give
\begin{equation}
\begin{split}
    \widetilde{\cA}_3^{\rm MHV} &= \prod_{i=1}^3\delta_{h_-\hb_i-1,0} \int_0^\infty d\omega_i \, \omega_i^{\Delta_i-1}  \cA_{3}^{\text{MHV}}\\
                  &= -4\prod_{i=1}^3\delta_{h_-\hb_i-1,0}\int_0^\infty d\omega_3 \,\omega_3^{(\sum_{j=1}^3 \Delta_j)-4} \delta(\zb_{31})\delta(\zb_{23}) z_{21}^{3-\Delta_1-\Delta_2} z_{23}^{\Delta_1-4} z_{31}^{\Delta_2-4} \times\\
                  &\delta^{(8)}\left( \sqrt{\frac{z_{23}}{z_{21}}}\xi_1\tau_1^A + \sqrt{\frac{z_{31}}{z_{21}}}\xi_2 \tau_2^A -\xi_3\tau_3^A \right)\, .
\end{split}
\end{equation}
To ensure that the conformal primary wavefunctions are delta-function normalisable we must have $\Delta_i = 1+ i\beta_i$ \cite{Pasterski:2016qvg}, hence the amplitude becomes 
\begin{align}
        \widetilde{\cA}_3^{\rm MHV} &= -(8\pi)\prod_{i=1}^3\delta_{h_i-\hb_i-1,0}\,\delta\left(\sum_{i=1}^3\beta_i\right) \delta(\zb_{31})\delta(\zb_{23}) z_{21}^{\Delta_3-4} z_{23}^{\Delta_1-4} z_{31}^{\Delta_2-4} \times\\ \nonumber
                  &\qquad\qquad \delta^{(8)}\left( \sqrt{z_{23}}\xi_1\tau_1^A + \sqrt{z_{31}}\xi_2 \tau_2^A -\sqrt{z_{21}}\xi_3\tau_3^A \right)\\ \nonumber
                 &= -(8\pi)\prod_{i=1}^3\delta_{h_i-\hb_i-1,0}\,\delta\left(\sum_{i=1}^3\beta_i\right) \delta(\zb_{31})\delta(\zb_{23})\prod_{i<j} z_{ij}^{3-2h_i-2h_j}
        \times\\ \nonumber & \qquad\qquad \delta^{(8)}\left( \Big(\frac{z_{23}}{z_{31}z_{12}}\Big)^{1/4} \xi_1\tau_1^A + \Big(\frac{z_{31}}{z_{12}z_{23}}\Big)^{1/4}\xi_2 \tau_2^A -\Big(\frac{z_{12}}{z_{23}z_{31}}\Big)^{1/4}\xi_3\tau_3^A \right)        \, ,
\end{align}
where in the last expression we have extracted the term carrying the conformal scaling properties for the variables $z$.
To check this result we can extract the three-point MHV gluon amplitude from the superamplitude by acting with the appropriate Gra{\ss}mann derivatives $\pld{A}= e^{(\pld{h}-\pld{\hb})/4}\check{\partial}_{A}$,
\begin{align}
    \widetilde{A}^{\rm{MHV}}_3(-,-,+)&= (\partial_{1A})^4(\partial_{2A})^4\tilde{\cA}^{\rm MHV}=(\check{\partial}_{1A})^4(\check{\partial}_{2A})^4 e^{(\pld{h_1}-\pld{\hb_1})} e^{(\pld{h_2}-\pld{\hb_2})} \tilde{\cA}^{\rm MHV}\\ \nonumber
    &=-(8\pi)\delta_{h_1-\hb_1+1,0}\delta_{h_2-\hb_2+1,0}\delta_{h_3-\hb_3-1,0}\,\delta\left(\sum_{i=1}^3\beta_i\right) \delta(\zb_{31})\delta(\zb_{23}) z_{21}^{\Delta_3} z_{23}^{\Delta_1-2} z_{31}^{\Delta_2-2}.
\end{align}
If we strip off the helicity constraint deltas from each leg this result agrees with \cite{Pasterski:2017ylz} up to an overall normalisation. 

We can follow a similar calculation for the four-point MHV superamplitude. As before, we rewrite the momentum-conserving delta function
\begin{equation}
    \begin{split}
        \delta^{(4)}(p_1+p_2-p_3-p_4)&= \delta^{(4)}(\omega_1\xi_1\tilde{\xi}_1 +\omega_2\xi_2\tilde{\xi}_2-\omega_3\xi_3\tilde{\xi}_3-\omega_4\xi_4\tilde{\xi}_4)\\
        &= \frac{4\delta(r-\bar{r})}{\omega_1 \abs{z_{13}}^2\abs{z_{24}}^2}
        \delta\left(\omega_4+\omega_1\frac{z_{13}\zb_{12}}{z_{34}\zb_{24}}\right) \delta\left(\omega_3-\omega_1\frac{z_{14}\zb_{12}}{z_{34}\zb_{23}}\right)
        \delta\left(\omega_2+\omega_1\frac{z_{13}\zb_{14}}{z_{23}\zb_{24}}\right) ,
    \end{split}
\end{equation}
where $r$ is the cross-ratio given by
\begin{equation}
    r = \frac{z_{12}z_{34}}{z_{13}z_{24}}\, .
\end{equation}
The $\delta(r-\bar{r})$ term imposes the constraint from $2\to 2$ scattering that all four particles lie in the same plane \cite{Arkani-Hamed:2020gyp}. Again we have three delta functions which trivialise the integrals over $\omega_{2,3,4}$ and the remaining integral over $\omega_1$ gives us a $\delta(\beta)$ factor, where $\beta = \sum_{i=1}^4\beta_i$. Thus the four-point celestial MHV superamplitude is
\begin{equation}
    \begin{split}
        \widetilde{\cA}_4^{\rm MHV}&= (8\pi)\prod_{i=1}^4\delta_{h_i-\hb_i-1,0}\,\delta(\beta) \left(\frac{z_{13}}{\zb_{42}}\right)^{\Delta_2+\Delta_4-4} \left(\frac{\zb_{12}}{z_{34}}\right)^{\Delta_3 + \Delta_4 -4} \left(\frac{\zb_{14}}{z_{23}}\right)^{\Delta_2-2} \left(\frac{z_{14}}{\zb_{23}}\right)^{\Delta_3-2}\times\\
        &\frac{\delta(r-\bar{r})}{z_{12}z_{23}z_{34}z_{41} \abs{z_{13}}^2\abs{z_{24}}^2} \delta^{(8)}\left( \xi_1\tau_1^A + \sqrt{\frac{ z_{13}\zb_{14}}{z_{23}\zb_{42}} }\xi_2 \tau_2^A -\sqrt{\frac{z_{14}\zb_{12}}{z_{34}\zb_{23}}}\xi_3\tau_3^A  - \sqrt{\frac{z_{13}\zb_{12}}{z_{34}\zb_{42}}}\xi_{4}\tau^4\right)\, .
    \end{split}
\end{equation}
Defining the conformally covariant expression
\begin{equation}
    I_4(z,\bar z) = \prod_{i<j}z_{ij}^{\frac{h}{3}-(h_i+h_j)}\bar z_{ij}^{\frac{\hb}{3}-(\bar h_i+\bar h_j)}\, , 
\end{equation}
where $h = \sum_{i=1}^4h_i$ and $\hb = \sum_{i=1}^4\hb_i$ and,
\begin{equation}
    x_{i}=\prod_{j<k}(z_{jk}\bar z_{jk})^{\frac{1}{12}}\prod_{l\not=i}(z_{il}\bar z_{il})^{-\frac{1}{4}}=I_4^{-1}e^{\frac{\partial_{h_i}+\partial_{\hb_i}}{4}}I_4\, ,
\end{equation}
then the above amplitude may be written as (with $\epsilon_1=\epsilon_2=-\epsilon_3=-\epsilon_4=1$)
\begin{equation}\label{bills4point}
    \begin{split}
        \widetilde{\cA}_4^{\rm MHV}&= (8\pi)\prod_{i=1}^4\delta_{h_i-\hb_i,1}\,\delta(\beta)\,\delta(r-\bar{r}) \,I_4(z,\bar z)\,
         r^{\frac{1}{3}}(1-r)^{\frac{1}{3}}\,
 \delta^{(8)}\left( \sum_{i=1}^4 \epsilon_i x_{i} \xi_i \tau_i^A\right)\, .
    \end{split}
\end{equation}
Now we can write
\begin{equation}
    \delta^8\left( \sum_{i=1}^4 \epsilon_i x_i \xi_{i\;\alpha} \tau_i^A\right)=2\prod_{A=1}^{4} \sum_{i< j} \epsilon_i \epsilon_j x_ix_j z_{ij} \tau_i^A\tau_j^A\, , 
\end{equation}
and conclude that this fermionic delta function is weightless under $(z_i,\tau_i^A)\rightarrow(z_i',\tau_i^{\prime A})$; this is only made possible using a supercoordinate $\tau^A$ which transforms under M\"obius maps.  The result \eqref{bills4point} for the superamplitude is consistent with the four-point gluon amplitude appearing in, for example, \cite{Pasterski:2017ylz}.

Expressions for the bosonic $n$-point amplitudes have been derived in \cite{Schreiber:2017jsr, Fan:2019emx}. At five points, we define the conformally covariant expression
\begin{equation}
    I_5(z,\bar z) = \prod_{i<j}z_{ij}^{\frac{h}{6}-\frac{2}{3}(h_i+h_j)}\bar z_{ij}^{\frac{\hb}{6}-\frac{2}{3}(\bar h_i+\bar h_j)}\, . 
\end{equation}
Then we find that the five-point MHV superamplitude can be written in the form
\begin{equation}\label{bills5point}
    \begin{split}
        \widetilde{\cA}_5^{\rm MHV}&\sim\prod_{i=1}^5\delta_{h_i-\hb_i,1}\,\delta(\beta) \,I_5(z,\bar z)\, X(r,\{h_i,\hb_i\})\,
 \delta^{(8)}\left( \sum_{i=1}^5   \epsilon_i y_i V_i(r)\,\xi_i \tau_i^A\right)\, ,
   \end{split}
\end{equation}
where
\begin{equation}
    y_i=\prod_{j<k}(z_{jk}\bar z_{jk})^{\frac{1}{24}}\prod_{l\not=i}(z_{il}\bar z_{il})^{-\frac{1}{6}}=I_5^{-1}e^{\frac{\partial_{h_i}+\partial_{\hb_i}}{4}} I_5\, ,
\end{equation}
and
\begin{equation}
    V_i(r)=X^{-1}e^{\frac{\partial_{h_i}+\partial_{\hb_i}}{4}} X\, .
\end{equation}
Here $X(r, \{h_i,\hb_i\})$ is a function only of the cross-ratios (of which there are two at five-point) and the weights $\{h_i,\hb_i\}$ which appear as exponents. The amplitude \eqref{bills5point} transforms covariantly under two-dimensional conformal transformations with the correct weights.

For $n>5$ points, the bosonic amplitudes are given in terms of $n-4$ integrals. In the formulation of \cite{Fan:2019emx}, one can show that there are changes of integration variables such that the $z,\zb$ dependence inside the integrals is only through cross-ratios. For the superamplitudes, this approach can also be followed, and the outcome is that the superamplitude is given by integrals over expressions which take a form analogous to the right-hand side of \eqref{bills5point}, where the function $X$ and the coefficients inside the supersymmetric delta function now depend on the integration variables and the cross-ratios. These expressions are rather complicated and it would be of interest to explore if there is a  more concise formulation.

\section{Celestial superconformal algebra} \label{sec:Celestial SC algebra}
 With our description of celestial on-shell superspace in chiral coordinates and a chiral Mellin transform taking us to Mellin space, we can derive the generators of superconformal symmetry. We will follow the same path as in Section \ref{sec: CC algebra} to write the generators on the celestial sphere in terms of their momentum space counterparts.

Our first step is to write the spinors $\ld{\alpha}$, $\ltd{\alpha}$ and super-coordinate $\eta^A$ in chiral coordinates,  and perform a change of variables from $\{\ld{\alpha},\ltd{\alpha},\eta^A\}$ to  $\{z,\zb,u,\ub,\tau^A\}$. Since $\eta^A=\left({\ub}/{u}\right)^{1/2}\tau^A$, the re-parameterisation of superspace in chiral coordinates will also change the form of the spinor derivatives, for example
\begin{equation}
\begin{split}
    \frac{\partial}{\partial \lambda^1}&= \partdiv{z}{\lu{1}}\partdiv{}{z} +  \partdiv{\zb}{\lu{1}}\partdiv{}{\zb} + \partdiv{u}{\lu{1}}\partdiv{}{u} +  \partdiv{\ub}{\lu{1}}\partdiv{}{\ub} +  \partdiv{\tau^A}{\lu{1}}\partdiv{}{\tau^A}\\
           &= -\frac{z}{u}\partdiv{}{z} + \partdiv{}{u} + \frac{1}{2}\frac{1}{u}\tau^A\partdiv{}{\tau^A}\, .
\end{split}
\end{equation}
We see that  $\tau^A$ now appears explicitly in the spinor derivatives, however this is not  surprising given the form of the Lorentz generators we found  in \eqref{eq: super Lorentz}. For the remaining objects we have
\begin{equation}\label{momentumspaceoperators}
 \begin{aligned}
    &\ld{\alpha}=u\begin{pmatrix} z\\1 \end{pmatrix}_{\alpha} , 
    &\qquad
    &\pld{\alpha}=\frac{1}{u}\begin{pmatrix}
        u\partial_u+ \frac{1}{2}\tau^A\check{\partial}_A-z\partial\\ -\partial
    \end{pmatrix}_{\alpha},\\
    &\ltd{\alpha}=u \begin{pmatrix} \zb\\1 \end{pmatrix}_{\dot{\alpha}} ,  &\qquad
    &\pltd{\alpha}=\frac{1}{\ub}\begin{pmatrix}
        \ub\partial_{\ub}-\frac{1}{2}\tau^A\check{\partial}_A-\zb\partialb,\\ -\partialb
    \end{pmatrix}_{\alpha},\\
    &\eta^A=\left(\frac{\ub}{u}\right)^{1/2}\tau^A,
    &\qquad
    &\partial_A=\left(\frac{\ub}{u}\right)^{-1/2}\check{\partial}_A.\\
\end{aligned}
\end{equation}
As in Section \ref{sec: CC algebra}, these are uniquely mapped to operators in Mellin space under the chiral Mellin transform. We obtain%
\footnote{Again we use the same expression for a momentum space operator and its Mellin space counterpart.}
\begin{equation}\label{eq: celest spinor eta gens}
 \begin{aligned}
    &\ld{\alpha}=\begin{pmatrix} z\\1 \end{pmatrix}_{\alpha}\,  e^{\frac{\partial_h}{2}}, 
    &\quad
    &\pld{\alpha}=-\begin{pmatrix}
        2(h-\frac{1}{4}\tau^A\check{\partial}_A)-1+z\partial\\ \partial
    \end{pmatrix}_{\alpha}\, e^{-\frac{\partial_h}{2}},\\
    &\ltd{\alpha}=\begin{pmatrix} \zb\\1 \end{pmatrix}_{\dot{\alpha}} e^{\frac{\partial_{\hb}}{2}},  &\qquad
    &\pltd{\alpha}=-\begin{pmatrix}
        2(\hb+\frac{1}{4}\tau^A\check{\partial}_A)-1+\zb\partialb,\\ \partialb
    \end{pmatrix}_{\alpha}e^{-\frac{\partial_{\hb}}{2}},\\
    &\eta^A=\tau^A\, e^{-\frac{\partial_h}{4}+\frac{\partial_{\hb}}{4}},
    &\qquad
    &\partial_A=\check{\partial}_A\, e^{\frac{\partial_h}{4}-\frac{\partial_{\hb}}{4}}.\\
\end{aligned}
\end{equation}
Note that $h$ and $\bar{h}$ are now shifted in the same way as in \eqref{hshift}.

From here we can write down the superconformal generators on the celestial sphere, using their representation in on-shell superspace (as in e.g.~\cite{Drummond:2008vq,Drummond:2009YS}) and the dictionary provided by~\eqref{eq: celest spinor eta gens}. 

First we have translations, Lorentz, conformal and dilation generators: 
\begin{align}\label{celestconformalgens}
    &p_{\alpha\dot{\alpha}}=\ld{\alpha}\ltd{\alpha}=
    \begin{pmatrix}
		z\zb & z \\
		\zb & 1 
	\end{pmatrix}_{\alpha\dot{\alpha}}\,  e^{\frac{\partial_h}{2}+\frac{\partial_{\hb}}{2}}\, , \\
	&m_{\alpha\beta}= \ld{(\alpha}\partial_{\beta)}=\begin{pmatrix}
         -2z(h-\frac{1}{4}\tau^A\check{\partial}_A) -z^2\partial & -(h-\frac{1}{4}\tau^A\check{\partial}_A)-z\partial\\ \cr
         -(h-\frac{1}{4}\tau^A\check{\partial}_A)-z\partial & -\partial
    \end{pmatrix}_{\alpha\beta}\, , \\
    &\bar{m}_{\dot{\alpha}\dot{\beta}}=\tilde{\lambda}_{(\dot{\alpha}}\pltd{\beta)}=
    \begin{pmatrix}
         -2\zb(\hb+\frac{1}{4}\tau^A\check{\partial}_A) -\zb^2\partialb & -(\hb+\frac{1}{4}\tau^A\check{\partial}_A)-\zb\partialb\\ \cr
         -(\hb+\frac{1}{4}\tau^A\check{\partial}_A)-\zb\partialb & -\partialb
    \end{pmatrix}_{\dot{\alpha}\dot{\beta}}\, , \\
    &k_{\alpha\dot{\alpha}}=\pld{\alpha}\pltd{\alpha}\nonumber\\
    &= \begin{pmatrix}
    (2(h-\frac{1}{4}\tau^A\check{\partial}_A)-1+z\partial)(2(\hb+\frac{1}{4}\tau^A\check{\partial}_A)-1+\bar{z}\bar{\partial}) & (2(h-\frac{1}{4}\tau^A\check{\partial}_A)-1+z\partial)\bar{\partial}\\ \cr
    (2(\hb+\frac{1}{4}\tau^A\check{\partial}_A)-1+\bar{z}\bar{\partial})\partial & \partial\bar{\partial}
    \end{pmatrix}_{\alpha\dot{\alpha}}\\ 
    &\hspace{115mm} \times\,  e^{-\frac{\partial_{h}}{2}-\frac{\partial_{\hb}}{2}}\, , \nonumber\\ \cr
    &d=\frac{1}{2}\lu{\alpha}\pld{\alpha}+ \frac{1}{2}\ltu{\alpha}\pltd{\alpha}+1=-(h-\frac{1}{4}\tau^A\check{\partial}_A+\hb+\frac{1}{4}\tau^A\check{\partial}_A-1)=-(h+\hb-1)\, .
\end{align}
The $\cN=4$ supersymmetry generators are given by:
\begin{align}\label{celestsusygens}
    q^{A}_{\alpha}= \ld{\alpha}\eta^A=
    \begin{pmatrix}
z\\1
\end{pmatrix}_{\alpha}\tau^{A}\,  e^{\frac{\partial_{h}}{4}+\frac{\partial_{\hb}}{4}} , \qquad \quad  \bar{q}_{\dot{\alpha}A} = \ltd{\alpha}\pld{A}= \begin{pmatrix} \bar{z}\\ 1
\end{pmatrix}_{\dot{\alpha}}\check{\partial}_{A} \, e^{\frac{\partial_{h}}{4}+\frac{\partial_{\hb}}{4}} \, .
\end{align}
This matches the structure of the $\cN=1$ supersymmetry generators found in \cite{Fotopoulos:2020bqj}. The remaining superconformal generators on the celestial sphere are
\begin{align}\label{celestsupconfgens}
    s_{\alpha A}&= \pld{\alpha} \pld{A}=
    -\begin{pmatrix}
        2(h-\frac{1}{4}\tau^A\check{\partial}_A)-1+z\partial\\
        \partial
    \end{pmatrix}_{\alpha} \check{\partial}_A\, e^{-\frac{\partial_{h}}{4}-\frac{\partial_{\hb}}{4}}\, , 
    \\
    \bar{s}_{\dot{\alpha}}^{A}&= \pltd{\alpha}\eta^A=
    -\begin{pmatrix}
        2(\hb+\frac{1}{4}\tau^A\check{\partial}_A)-1+\bar{z}\bar{\partial}\\\bar{\partial}
    \end{pmatrix}_{\dot{\alpha}}\tau^{A} \, e^{-\frac{\partial_h}{4}-\frac{\partial_{\hb}}{4}}\, , \\
    r^A_B&=\tau^A\check{\partial}_B-\frac{1}{4}\delta^A_B \tau^C \check{\partial}_C\, , \\
    c &:=1-\mathcal{J}=1-(h-\frac{1}{4}\tau^A\check{\partial}_A-(\hb+\frac{1}{4}\tau^A\check{\partial}_A)+\frac{1}{2}\tau^A\check{\partial}_A)\nonumber\\
    \label{endofgens}
    &\, =1-(h-\hb)\, .
\end{align}
One can deduce that these generators satisfy the full superconformal algebra 
by checking that the commutation relations of the celestial superspace operator representation of the variables $\{\ld{\alpha},\ltd{\alpha},\eta^A\}$ and the derivatives with respect to them reproduce those of the superspace variables and derivatives.
These celestial superconformal generators are guaranteed to annihilate superamplitudes from their derivation, as was the case for the conformal generators in Section~\ref{sec: CC algebra}. 


\section{Conclusions}
\label{conclusions}

In this paper we  considered a celestial space parameterisation of null momenta in which the little group is not fixed. This leads to a natural definition of a ``chiral'' Mellin transform which can be reduced to the usual ``non-chiral'' transform, but which also includes a spin constraint $h - \bar{h} = J$.  We discussed how states and amplitudes are represented in these chiral celestial coordinates and  how this approach simplifies the derivation of the conformal symmetry operators. We extended this discussion to $\cN\!=\!4$ SYM, giving a definition of $\cN\!=\!4$ chiral superspace, the states and superamplitudes, and presented  some explicit examples. We further showed how the full superconformal algebra generators follow directly from their spacetime representations.

This work provides a potentially simplified approach to the study of (chiral) celestial amplitudes and their symmetries. It can be directly applied to the derivation of particular amplitudes and the formulation of recursion relations. Several avenues for further investigation are possible. Among these, one pressing question is to clarify the nature of a potential celestial  two-dimensional superconformal field theory whose correlation functions are our superamplitudes of $\cN\!=\!4$ SYM. Extensions to supergravity, in particular $\cN\!=\!8$ supergravity should  also be immediate, 
as should the study of amplitudes in theories with less than maximal supersymmetry, for which the appropriate on-shell superspace is known. It would also be very interesting to intensify the study of  celestial loop amplitudes, and to find a possible celestial representation of hidden symmetries of the $\cN\!=\!4$ SYM $S$-matrix such as the dual superconformal and Yangian symmetries. 
We will return to these questions in the near future.

\section*{Acknowledgements}

We would like to thank Rashid Alawadhi, Stefano De Angelis, Jung-Wook Kim,  Adrian Keyo Shan Padellaro and Congkao Wen for interesting discussions.  We also thank Hongliang Jiang for sharing his draft \cite{Jiang:2021xzy} with us, and Congkao Wen for  informing each of us about this parallel work. 
This work  was supported by the Science and Technology Facilities Council (STFC) Consolidated Grants ST/P000754/1 \textit{``String theory, gauge theory \& duality''} and  ST/T000686/1 \textit{``Amplitudes, strings  \& duality''}
and by the European Union's Horizon 2020 research and innovation programme under the Marie Sk\l{}odowska-Curie grant agreement No.~764850 {\it ``\href{https://sagex.org}{SAGEX}''}.
The work of GRB and JG  is supported by an STFC quota studentship. 

\appendix   
\section{Conformal weights}
\label{app:A}

In this appendix we state the weights for all objects appearing in $\mathcal{N}\!=\!4$ SYM, Table \ref{tab:tableofweights}, and provide some example derivations. The weights of an object are defined under the two-dimensional conformal group only once it has been mapped to Mellin space by a chiral Mellin transform, and are derived according to the method laid out in Section \ref{sec: weights}.

To summarise, the weights of generators are always dictated by their helicity and carry zero conformal dimension -- whereas splitting generators into their constituent differential operators produces objects with conformal dimension also. The weights of a celestial amplitude are given by the parameters in the chiral Mellin transform which have values $(h,\hb)=(\frac{\Delta+J}{2}, \frac{\Delta-J}{2})$. Intuitively, the spin $J$ is set by the amplitude itself while the conformal dimension $\Delta := h+\hb$ is a free parameter set by the powers of $\omega$ (equivalently $u,\ub$) in the Mellin transform. The weights of all operators are defined according to how they act on celestial amplitudes.

As examples we will now derive the weights of the Gra{\ss}mann variable $\eta^A$, the spinor derivative $\pld{\alpha}$ and the exponential operator $\, e^{\frac{\partial_h}{2}}$.

\begin{table}[ht]
    \centering
    \begin{tabular}{ |p{5cm}||p{3cm}|p{3cm}|  }
    \hline
     \multicolumn{3}{|c|}{\bf Table of conformal weights} \\
    \hline
    Celestial Object & $(h,\hb)$ & $(\Delta,J)$\\
    \hline
    &&\\
    $\widetilde{A}_J$   &  $(\frac{\Delta+J}{2}, \frac{\Delta-J}{2})$    &$(\Delta,J)$\\[5pt]
    $\ld{\alpha}$   &  $(-\frac{1}{4}, \frac{1}{4})$    &$(0,-\frac{1}{2})$\\[5pt]
    $\ltd{\alpha}$   &  $(\frac{1}{4}, -\frac{1}{4})$    &$(0,\frac{1}{2})$\\[5pt]
    $\pld{\alpha}$   &  $(\frac{1}{4}, -\frac{1}{4})$    &$(0,\frac{1}{2})$\\[5pt]
    $ \pltd{\alpha}$   &  $(-\frac{1}{4}, \frac{1}{4})$    &$(0,-\frac{1}{2})$\\[5pt]
    $\eta^A$   &  $(\frac{1}{4}, -\frac{1}{4})$    &$(0,\frac{1}{2})$\\[5pt]
    $\partial_A$   &  $(-\frac{1}{4}, \frac{1}{4})$    &$(0,-\frac{1}{2})$\\[5pt]
     $\xi_{\alpha}=(z,1)_{\alpha}$   &  $(-\frac{1}{2}, 0)$    &$(-\frac{1}{2},-\frac{1}{2})$\\[5pt]
     $\tilde{\xi}_{\dot\alpha}=(\zb,1)_{\dot\alpha}$   &  $(0, -\frac{1}{2})$    &$(-\frac{1}{2},\frac{1}{2})$\\[5pt]
     $\begin{pmatrix}
        -2(h-\frac{1}{4}\tau^A\check{\partial}_A)+1-z\partial\\-\partial\end{pmatrix}_{\alpha}$   &  $(\frac{1}{2}, 0)$    &$(\frac{1}{2},\frac{1}{2})$\\[5pt]
    $\begin{pmatrix}
        -2(\hb+\frac{1}{4}\tau^A\check{\partial}_A)+1-\zb\partialb\\-\partialb\end{pmatrix}_{\dot\alpha}$   &  $(0, \frac{1}{2})$    &$(\frac{1}{2},-\frac{1}{2})$\\[5pt]
    $\tau^A$   &  $(\frac{1}{4}, -\frac{1}{4})$    &$(0,\frac{1}{2})$\\[5pt]
    $\check{\partial}_A$   &  $(-\frac{1}{4}, \frac{1}{4})$    &$(0,-\frac{1}{2})$\\[5pt]
    $e^{\frac{\partial_h}{2}}$   &  $(\frac{1}{4}, \frac{1}{4})$    &$(\frac{1}{2},0)$\\[5pt]
    $\, e^{\frac{\partial_{\hb}}{2}}$   &  $(\frac{1}{4}, \frac{1}{4})$    &$(\frac{1}{2},0)$\\[5pt]
    $\, e^{-\frac{\partial_h}{4}+\frac{\partial_{\hb}}{4}}$   &  $(0, 0)$    &$(0,0)$\\[5pt]
    \hline
    \end{tabular}
    \caption{Table of conformal weights for objects defined or acting in Mellin space.}
    \label{tab:tableofweights}
\end{table}


\textbf{Gra{\ss}mann variable: } $\eta^A$ 

Following Section \ref{sec: weights} we can compute the weights of the celestial operator $\eta^A=\tau^A\, e^{-\frac{\partial_h}{4}+\frac{\partial_{\hb}}{4}}$ by acting on a celestial superstate $|z,\zb,h,\hb,\tau^A\rangle$. Under a Lorentz transformation 
\begin{align}
    \eta^A\, |z,\zb,h,\hb,\tau^A\rangle\rightarrow(cz+d)^{-2h}(\cb\zb+\db)^{-2\hb}\eta^A|z,\zb,h,\hb,\tau^A\rangle\, .
\end{align}
In order to find the weight values that $h,\hb$ localise onto, we look at the action of $\eta^A$ (as given above) on the Kronecker delta. This gives the following shifted spin constraints:
\begin{align}
    &h-\hb=1+\frac{1}{2}, \qquad h+\hb=\Delta=1+i\lambda\, , 
    \end{align}
    hence
    \begin{align}
    h=\frac{\Delta+1}{2}+\frac{1}{4}\, , \qquad \hb=\frac{\Delta-1}{2}-\frac{1}{4}\, .
\end{align}
We see that the operator $\eta^A$ has shifted the weights by $(\frac{1}{4},-\frac{1}{4})$ in line with an object with helicity $\frac{1}{2}$, while identical considerations assigns weights $(-\frac{1}{4},\frac{1}{4})$ to $\partial_A=\check{\partial}_A\, e^{\frac{\partial_h}{4}-\frac{\partial_{\hb}}{4}}$. \\

\textbf{Spinor Derivative: } $\pld{\alpha}$

The spinor derivative in the supersymmetric theory is written in momentum space according to \eqref{momentumspaceoperators} and transforms under a Lorentz transformation as
\begin{align}\label{spinorderivativelorentzmap}
&M_{\alpha}^{\ \beta}\, \partial_{\beta}\, =\, \begin{pmatrix}
a & b\\
c & d
\end{pmatrix}_{\alpha}^{\ \beta}\partial_{\beta}=\frac{1}{u}\bigg[\begin{pmatrix}
a(u \partial_{u}+\frac{1}{2}\tau^A\check{\partial}_A)\\c(u \partial_{u}+\frac{1}{2}\tau^A\check{\partial}_A)\end{pmatrix}_{\alpha} - \begin{pmatrix}az+b \\
cz+d
\end{pmatrix}_{\alpha}\partial\bigg].
\end{align}
Under the combined M\"obius coordinate transformation $(u',\ub',z',\zb',\tau^{\prime\; A})$ we have the following partial derivatives,
\begin{equation}\label{partialderivs}
\begin{split}
    \partial_{u}&=(cz+d)\partial_{u^\prime}, \qquad \partial_{\ub}=(\cb\zb+\db)\partial_{\ub'}\, , \qquad \check{\partial}_A=\sqrt{\frac{cz+d}{\cb\zb+\db}}\check{\partial}^{\;'}_A\, ,\\
    \partial&=\frac{1}{(cz+d)^2}\partial'+\frac{c}{cz+d}u'\partial_{u'}+\frac{c}{2(cz+d)}\tau^{\prime A}\check{\partial}^{\;'}_A\, ,\\
    \partialb&=\frac{1}{(\cb\zb+\db)^2}\partialb'+\frac{\cb}{\cb\zb+\db}\ub'\partial_{\ub'}-\frac{\cb}{2(\cb\zb+\db)}\tau^{\prime A}\check{\partial}^{\;'}_A\, .
    \end{split}
\end{equation}
Using \eqref{partialderivs} we can write the Lorentz transformed spinor derivative after a few steps as,
\begin{align}\label{lorentsonspinorderiv}
M_{\alpha}^{\ \beta}\, \partial_{\beta}=\frac{1}{u'}\bigg[\begin{pmatrix}
    (u' \partial_{u'}+\frac{1}{2}\tau^{\prime A}\check{\partial}^{\;'}_A)\\0\end{pmatrix}_{\alpha} - \begin{pmatrix}z' \\
1
\end{pmatrix}_{\alpha}\partial'\bigg]\, ,
\end{align}
and, as was the case for the chiral spinors in \eqref{chiralsl2Clambda}, we have no need for additional phase factors.

Following Section \ref{sec: weights}, we can find the weights of the spinor derivative $\pld{\alpha}$ in Mellin space by acting on a superstate,
\begin{equation}
    \begin{split}
        \partial_{\alpha}|z,\zb,h,\hb,\tau^A\rangle=\int_{\mathbb{C}}\!du \wedge d\ub\,  u^{2h-1}\ub^{2\hb-1}\frac{1}{u}\bigg[\begin{pmatrix}
    u \partial_{u}+\frac{1}{2}\tau^{A}\check{\partial}_A\\0\end{pmatrix}_{\alpha} - \begin{pmatrix}z \\
    1
    \end{pmatrix}_{\alpha}\partial\bigg]|z',\zb',u',\ub',\tau^{\prime A}\rangle\, ,
    \end{split}
\end{equation}
and using \eqref{lorentsonspinorderiv} we conclude that this transforms under Lorentz with weight factors equal to $(cz+d)^{-2h}(\cb\zb+\db)^{-2\hb}$. However, the Kronecker delta appearing in $\pld{\alpha}\widetilde{\cA}$ has a shifted spin constraint and the state has a reduced conformal dimension due to the action of $\, e^{-\partial_h/2}$. Hence,
\begin{align}
    h-\hb=1+\frac{1}{2}\, , \qquad h+\hb-\frac{1}{2}=\Delta-\frac{1}{2}=1+i\lambda-\frac{1}{2}
    \, , \end{align}
implying,
    \begin{align}h=\frac{\Delta+1}{2}+\frac{1}{4}\, , \qquad \hb=\frac{\Delta-1}{2}-\frac{1}{4}\, .
\end{align}
Thus the operator $\pld{\alpha}$ has shifted the weights by $(\frac{1}{4},-\frac{1}{4})$ and we assign it weights accordingly and in line with an object with helicity $+\frac{1}{2}$.\\

\textbf{Exponential Shift Operator: } $\, e^{\frac{\partial_h}{2}}$

We can also break up generators into their constituent differential operators and study the separate action of those on celestial amplitudes and derive their weights. As an  example, consider the action of the exponential operator $e^{\frac{\partial_h}{2}}$ on a state. It shifts $(h,\hb) \rightarrow (h',\hb')$, where we have $h'=h+\frac{1}{2}$ while $\hb'=\hb$ is left unchanged. Hence,
\begin{equation}\label{expoperatoraction-bis}
\begin{split}
     e^{\frac{\partial_h}{2}}|z,\zb,h,\hb;J\rangle &= \
     |z,\zb,h',\hb';J\rangle\\&=
    \delta_{h' - \bar{h}' -J, 0}\,   |z,\zb,\Delta';J\rangle \, ,
\end{split}
\end{equation}
where the conformal dimension has been shifted to $\Delta'=\Delta+\frac{1}{2}$ from the conformal dimension of the original state, $\Delta=1+i\beta$. Hence, the weights satisfy
\begin{align}
    h'-\bar{h}' =J\, , \qquad \quad h' + \bar{h}' = \Delta + \frac{1}{2}\, , 
\end{align}
which implies
\begin{align}\label{hprime}
    h' = \frac{\Delta+J}{2} + \frac{1}{4}\, , \qquad \quad \bar{h}' = \frac{\Delta- J}{2} + \frac{1}{4}\, . 
\end{align}
Hence, under a Lorentz transformation 
\begin{equation}\label{expoperatoraction}
\begin{split}
    \, e^{\frac{\partial_h}{2}}|z,\zb,h,\hb;J\rangle
    &\longrightarrow (cz+d)^{-2h'}(\cb\zb+\db)^{-2\hb'}\,\;e^{\frac{\partial_h}{2}}|z',\zb',h,\hb;J\rangle\\
    &= (cz+d)^{-\Delta-J-\frac{1}{2}}(\cb\zb+\db)^{-\Delta+J - \frac{1}{2}}\,\; e^{\frac{\partial_h}{2}}|z',\zb',h,\hb;J\rangle\, ,
\end{split}
\end{equation}
where in the last line we have used \eqref{hprime}. 
The quantity in  \eqref{expoperatoraction} transforms with weights 
$(\frac{\Delta+J}{2}+\frac{1}{4}, \frac{\Delta-J}{2}+\frac{1}{4}) $, and comparing to the weights  $(\frac{\Delta+J}{2}, \frac{\Delta-J}{2})$ of the original state we see that 
the operator $e^{\frac{\partial_h}{2}}$ has shifted them by
$(\frac{1}{4}, \frac{1}{4})$. Hence we conclude that this operator has weights $e^{\frac{\partial_h}{2}} \sim (\frac{1}{4}, \frac{1}{4})$. We stress that the operator $e^{\frac{\partial_h}{2}}$ receives these weights despite shifting $h \rightarrow h'=h+\frac{1}{2}$; this is because the spin constraint still requires that $h'-\hb'=J$ and therefore the new weights $h',\hb'$ must correspond to a spin-$J$ object. In other words, the conformal dimension $\Delta$ is the only free parameter of a celestial amplitude and so only it is shifted.


\newpage

\bibliographystyle{utphys}
\bibliography{remainder}

\end{document}